\documentclass[reprint,twocolumn]{revtex4-2}
\usepackage{amsfonts}
\usepackage{amsmath}
\usepackage{amssymb}
\usepackage{charter}
\usepackage{graphicx}
\usepackage{float}
\usepackage{amsmath}
\usepackage{amssymb}
\usepackage{charter}
\usepackage{graphicx}
\usepackage{subfigure}
\usepackage{graphicx}
\usepackage{epstopdf}
\usepackage{color}
\usepackage{tocvsec2}
\usepackage{enumerate}
\usepackage{graphicx}
\usepackage{dcolumn}
\usepackage{bm}

\begin{document}

\title{Phase estimation via coherent and photon-catalyzed squeezed vacuum
states }
\author{Zekun Zhao$^{1}$}
\author{Qingqian Kang$^{1,2}$}
\author{Huan Zhang$^{3}$}
\author{Teng Zhao$^{1}$}
\author{Cunjin Liu$^{1}$}
\author{Liyun Hu$^{1,4}$}
\thanks{hlyun@jxnu.edu.cn}
\affiliation{$^{{\small 1}}$\textit{Center for Quantum Science and Technology, Jiangxi
Normal University, Nanchang 330022, China}\\
$^{{\small 2}}$\textit{Department of Physics, Jiangxi Normal University
Science and Technology College, Nanchang 330022, China}\\
$^{{\small 3}}$\textit{School of Physics, Sun Yat-sen University, Guangzhou
510275, China}\\
$^{{\small 4}}$\textit{Institute for Military-Civilian Integration of
Jiangxi Province, Nanchang 330200, China}}

\begin{abstract}
The research focused on enhancing the measurement accuracy through the use
of non-Gaussian states has garnered increasing attention. In this study, we
propose a scheme to input the coherent state mixed with photon-catalyzed
squeezed vacuum state into the Mach-Zender interferometer to enhance phase
measurement accuracy. The findings demonstrate that photon catalysis,
particularly multi-photon catalysis, can effectively improve the phase
sensitivity of parity detection and the quantum Fisher information.
Moreover, the situation of photon losses in practical measurement was
studied. The results indicate that external dissipation has a greater
influence on phase sensitivity than the internal dissipation. Compared to
input coherent state mixed with squeezed vacuum state, the utilization of
coherent state mixed photon-catalyzed squeezed vacuum state, particularly
the mixed multi-photon catalyzed squeezed vacuum state as input, can enhance
the phase sensitivity and quantum Fisher information. Furthermore, the phase
measurement accuracy can exceed the standard quantum limit, and even surpass
the Heisenberg limit. This research is expected to significantly contribute
to quantum precision measurement.

\textbf{PACS: }03.67.-a, 05.30.-d, 42.50,Dv, 03.65.Wj
\end{abstract}

\maketitle

\section{Introduction}

With the advancement of science and technology, quantum metrology, which
utilizes non-classical quantum states to achieve high-precision measurement,
has been extensively applied \cite{1,2,3,4,5,6,7,8,9,10}. The Mach-Zehnder
interferometer (MZI) is widely used as a measurement instrument in quantum
precision measurement \cite{11,12,13,14}. In 1981, Caves proposed a scheme
involving the mixture of coherent state (CS) and squeezed vacuum state (SVS)
as the input state of MZI and found that the measurement accuracy can exceed
the standard quantum limit (SQL) $\Delta \varphi =1/\sqrt{\bar{N}}$, where $%
\bar{N}$ represents the total average photon number of the input state \cite%
{15}. Subsequently, the three main parts of quantum precision measurement,
namely, the preparation of input states, the interaction between the input
state and the system under consideration, and the detection of the output
state \cite{16,17}, have gained significant attention. Currently, numerous
quantum states such as the NOON state, twin-Fock state, and two-mode
squeezed vacuum (TMSV) state inputs have been investigated and proven to
surpass the SQL and even achieve the Heisenberg limit (HL) $\Delta \varphi
=1/\bar{N}$ \cite{18,19,20,21,22,23,24,25,26,27,28,29,30,31,32}. In terms of
the detection method, parity detection offers significant advantages in
improving measurement accuracy. In 2011, Seshadreesan \emph{et al.}
demonstrated that by inputting the CS mixed with SVS into MZI and utilizing
parity detection, the phase sensitivity can reach the HL \cite{33}.

Recent studies have shown that enhancing the non-classical properties of
quantum states through non-Gaussian operations is beneficial for improving
measurement accuracy \cite{34,35,36,37,38,39,40,41,42,43,44}. Non-Gaussian
operations such as photon subtraction, addition, and catalysis can be
simulated with optical beam splitters. Jia \emph{et al}. conducted a
detailed study of the non-classical properties of the non-Gaussian state
prepared through such non-Gaussian operations on the SVS \cite{45}. Their
findings indicate that photon-subtracted, photon-added, and photon-catalyzed
non-Gaussian operations can effectively enhance the non-classical properties
of quantum states. Utilizing coherent state mixed with photon-subtracted
(-added) SVS (PSSVS/PASVS) as input in MZI for quantum precision measurement
has demonstrated effective enhancements \cite{38,39}. Furthermore, given the
inevitable photon losses during actual measurement, increasing attention has
been devoted to understanding the impact of photon losses on measurement
accuracy, which holds great significance for quantum precision measurement
applications. Building on the insights from previous research, we propose
employing the CS mixed with photon-catalyzed SVS (PCSVS) as the input state
of MZI. We consider the phase sensitivity and quantum Fisher information
(QFI) under ideal and photon-loss conditions to develop an improved phase
measurement accuracy scheme. It is demonstrate that our scheme can
significantly enhance phase estimation, particularly in the context of
multi-photon catalysis, leading to measurement accuracy exceeding the SQL or
even reaching the HL.

The organization of this paper is as follows: In Sec. II, we initially
introduce the preparation process and fundamental properties of the PCSVS.
In Sec. III, we investigate the phase sensitivity with parity measurement
and the QFI in ideal circumstances. Sec. IV further examines the impact of
photon losses on the phase sensitivity, including external and internal
dissipations. In Sec. V, we explore the influence of photon losses on the
QFI. The last section provides a summary of the main results.

\section{Photon-catalyzed squeezed vacuum state}

Fig. 1 illustrates the input of the SVS in mode $b$ and the Fock state $%
\left \vert m\right \rangle _{c}$ in mode $c$ into the two ports of the
beamsplitter (BS0) with transmissivity $\eta $. Subsequently, the photon
detector at the output port of mode $c$ detects the Fock state $\left \vert
m\right \rangle _{c}$ with the identical photon number $m$, thereby leading
to the successful preparation of the PCSVS at the other output port of mode $%
b$. According to Ref. \cite{45}, the equivalent operator for the $m$-photon\
catalysis operation can be defined as

\begin{eqnarray}
\hat{O}_{m} &=&\left. _{c}\left \langle m\right \vert B\left( \eta \right)
\left \vert m\right \rangle _{c}\right.  \notag \\
&=&\left. \frac{\left( \sqrt{\eta }\right) ^{b^{\dagger }b+m}}{m!}\frac{%
\partial ^{m}}{\partial \tau ^{m}}\frac{1}{1-\tau }\left( \frac{1-\tau /\eta
}{1-\tau }\right) ^{b^{\dagger }b}\right \vert _{\tau =0},  \label{1}
\end{eqnarray}%
where $B\left( \eta \right) =\exp \left[ \left( b^{\dagger }c-bc^{\dagger
}\right) \arccos \sqrt{\eta }\right] $ is the BS0 operator. Consequently,
the expression of the PCSVS can be derived as

\begin{eqnarray}
\left \vert \psi _{PCSVS}\right \rangle _{b} &=&\frac{\hat{O}_{m}}{\sqrt{%
P_{m}}}S\left( r\right) \left \vert 0\right \rangle _{b}  \notag \\
&=&\left. \frac{W_{0}}{\sqrt{P_{m}}}\frac{\partial ^{m}}{\partial \tau ^{m}}%
\frac{\exp \left( Wb^{\dagger 2}\right) }{1-\tau }\right \vert _{\tau
=0}\left \vert 0\right \rangle _{b},  \label{2}
\end{eqnarray}%
where $W_{0}=\frac{\sqrt{\eta }^{m}}{m!\sqrt{\cosh r}}$, $W=-\frac{\eta
\left( 1-\tau /\eta \right) ^{2}\tanh r}{2\left( 1-\tau \right) ^{2}}$, $%
S\left( r\right) \left \vert 0\right \rangle _{b}=\exp \left[ r\left(
b^{2}-b^{\dagger 2}\right) /2\right] \left \vert 0\right \rangle _{b}$ is
the SVS with the squeezing parameter $r$ and $P_{m}$ represents a normalized
coefficient, i.e.,

\begin{equation}
P_{m}=\left. W_{0}^{2}\frac{\partial ^{2m}}{\partial \tau ^{m}\partial \tau
_{1}^{m}}\epsilon \left( 1-4W_{1}W\right) ^{-\frac{1}{2}}\right \vert _{\tau
=\tau _{1}=0},  \label{3}
\end{equation}%
and

\begin{eqnarray}
W_{1} &=&-\frac{\eta \left( 1-\tau _{1}/\eta \right) ^{2}\tanh r}{2\left(
1-\tau _{1}\right) ^{2}},  \notag \\
\epsilon &=&\frac{1}{\left( 1-\tau \right) \left( 1-\tau _{1}\right) }.
\label{4}
\end{eqnarray}

\begin{figure}[tph]
\label{Fig1} \centering \includegraphics[width=0.83\columnwidth]{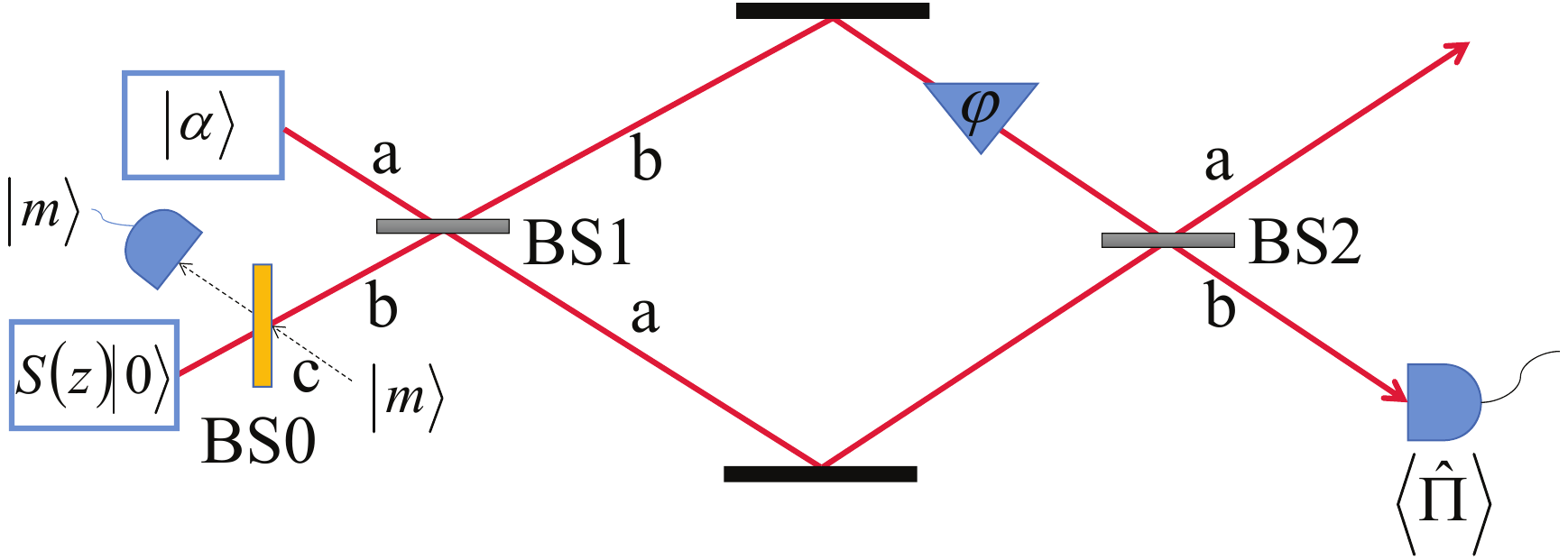}
\caption{The schematic diagram of the preparation of PCSVS and the parity
detection of phase shift when the PCSVS and CS are injected from the two
ports of the first beamsplitter BS1 of the MZI under ideal conditions.}
\end{figure}

The average photon number of an input state, which reflects the energy
characteristics of a light field, is a crucial parameter in the
investigation of quantum precision measurement. Therefore, we further
examine the average photon number of the PCSVS. Using Eq. (\ref{2}), the
average photon number of PCSVS is given by

\begin{eqnarray}
\overline{n}_{b} &=&\left. _{b}\left \langle \psi _{PCSVS}\right \vert
b^{\dagger }b\left \vert \psi _{PCSVS}\right \rangle _{b}\right.  \notag \\
&=&\hat{D}\frac{\epsilon }{\left( 1-4W_{1}W\right) ^{\frac{3}{2}}}-1,
\label{5}
\end{eqnarray}%
where

\begin{equation}
\hat{D}=\frac{W_{0}^{2}}{P_{m}}\frac{\partial ^{2m}}{\partial \tau
^{m}\partial \tau _{1}^{m}}\left \{ \cdot \right \} |_{\tau =\tau _{1}=0}.\
\label{6}
\end{equation}

\begin{figure}[tbh]
\label{Fig2} \centering
\subfigure{
\begin{minipage}[b]{0.5\textwidth}
\includegraphics[width=0.83\textwidth]{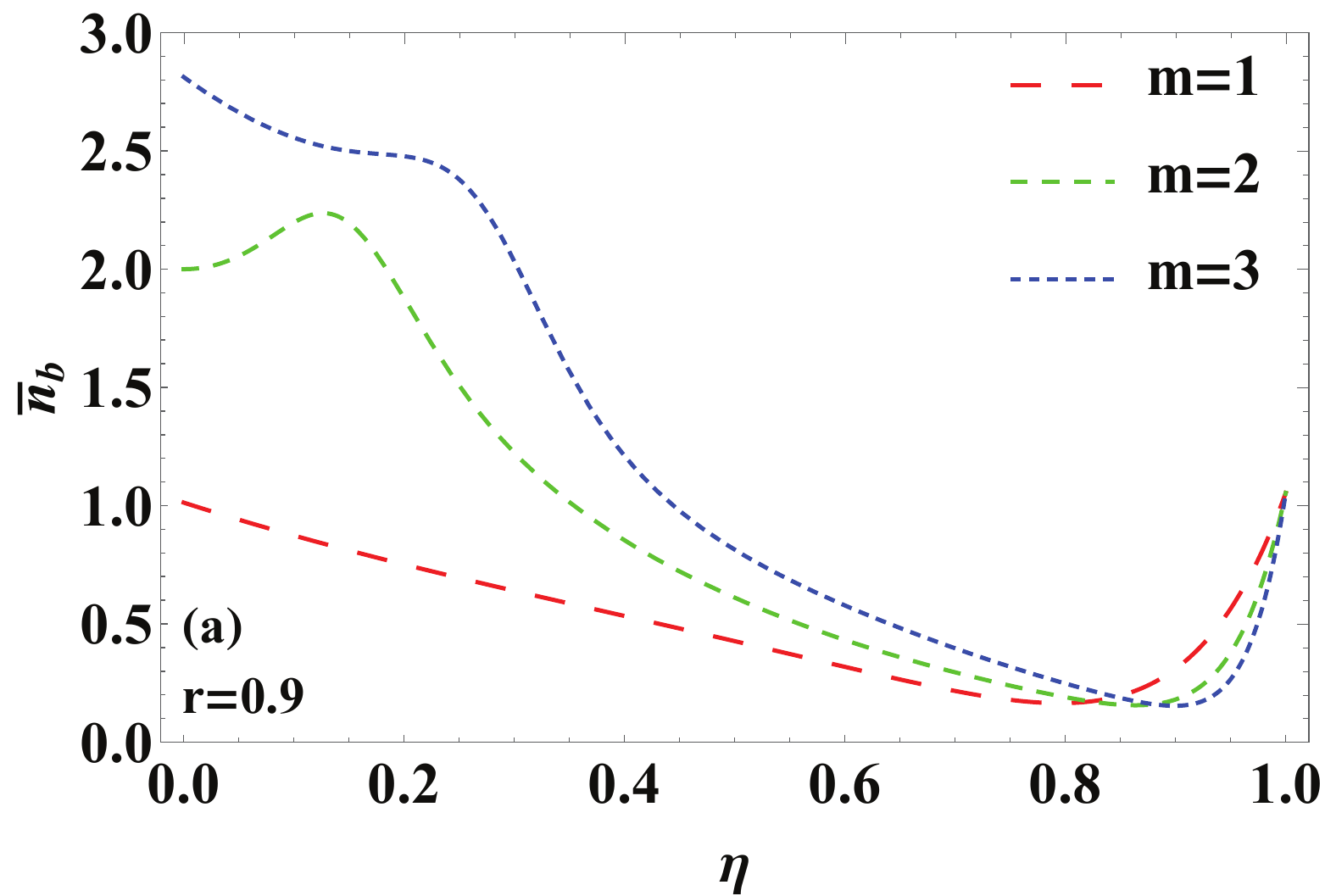}\\
\includegraphics[width=0.83\textwidth]{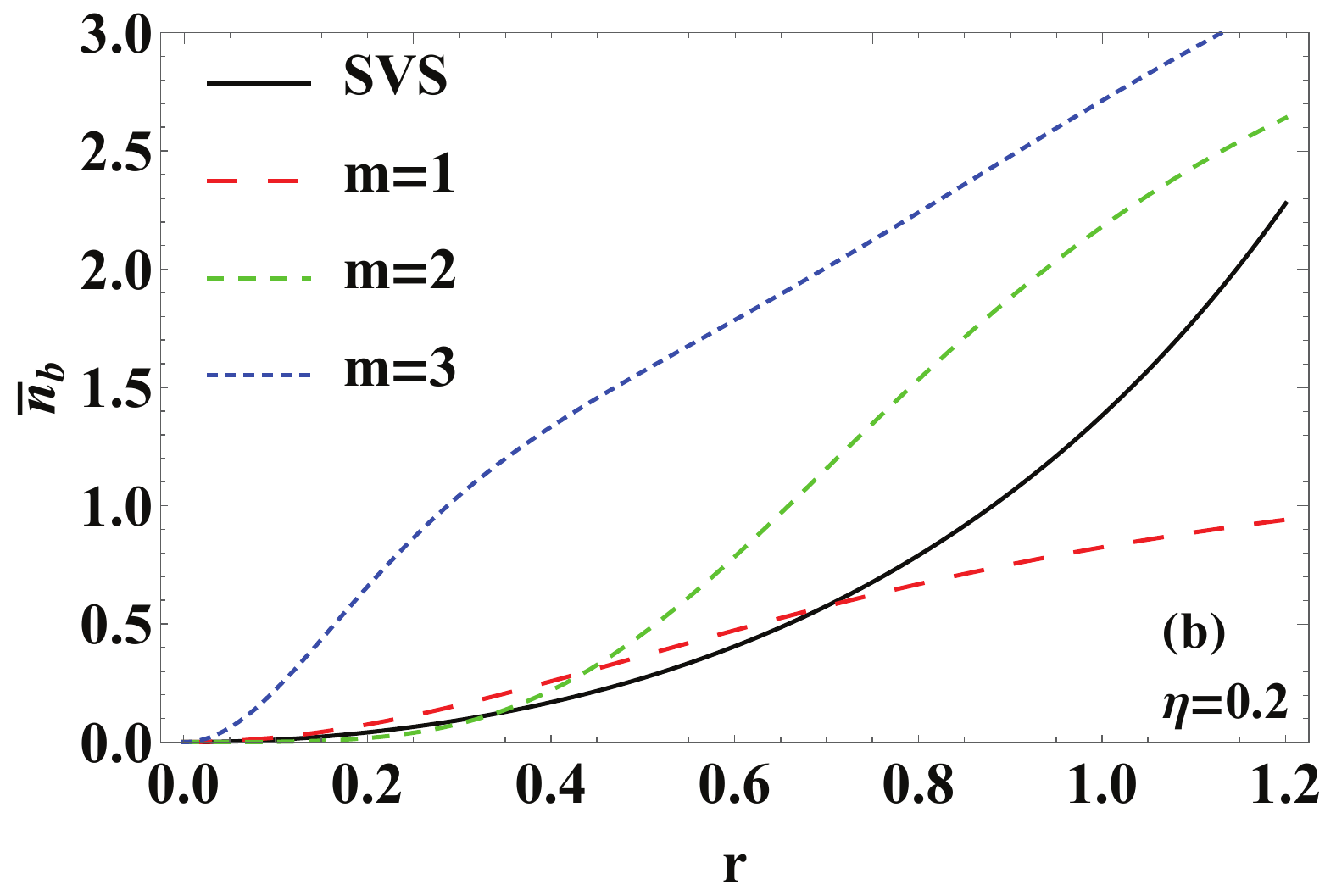}
\end{minipage}}
\caption{For the catalysis photon numbers $m=1,2,3$, (a) the average photon
number $\overline{n}_{b}$ as a function of the transmissivity $\protect \eta $
for a given squeezing parameter $r=0.9$, and (b) $\overline{n}_{b}$ as a
function of $r$ for a given $\protect \eta =0.2$ and its comparison with the
case of SVS.}
\end{figure}

In order to comprehensively analyze the relationship between the average
photon number and each parameter, for different catalysis photon numbers $%
m=1,2,3$, Fig. 2 illustrates the variation of the average photon number $%
\overline{n}_{b}$ of PCSVS with the transmissivity $\eta $ at a fixed
squeezing parameter $r=0.9$ (Fig. 2(a)), and the variation of $\overline{n}%
_{b}$ with $r$ for a given $\eta =0.2$ (Fig. 2(b)). From Fig. 2(a), it is
observed that $\overline{n}_{b}$ increases with an increase of the catalysis
photon number $m$ when the transmissivity $\eta $ is in the range of $0$ to
approximately $0.85$ for a fixed $r=0.9$, whereas $\overline{n}_{b}$
increases with a decrease of $m$ when $\eta $ is in the range of about $0.85$
to $1$. The $\overline{n}_{b}$ corresponding to $\eta =1$ is equal to the
average photon number of SVS. A comparison indicates that the $\overline{n}%
_{b}$ of the multiphoton catalysis SVS (MC-SVS) corresponding to $m\geqslant
2$ is greater than the average photon number of SVS when $\eta $ takes a
smaller value. Fig. 2(b) presents the plot of the average photon number of
the SVS and the PCSVS at a specific transmissivity $\eta =0.2$ as a function
of the squeezing parameter $r$. As demonstrated in Fig. 2(b), at $\eta =0.2$%
, $\overline{n}_{b}$ rises with $r$, and the $\overline{n}_{b}$ of MC-SVS
surpasses that of SVS within a specific range of $r$. Notably, in cases
where the squeezing parameter $r$\ is relatively small, the average photon
number $\overline{n}_{b}$ of the single-photon PCSVS ($m=1$) can also exceed
that of SVS.

\section{Phase estimation of mixing PCSVS with CS based on MZI under ideal
conditions}

\subsection{Phase sensitivity with parity detection}

The conventional MZI model, depicted in Fig. 1, primarily comprises two
50:50 beam splitters (BS1 and BS2) with equal transmissivity and
reflectivity, two mirrors, and a phase shifter. The beam splitter's impact
on the input state is equivalent to a rotation in theory. Therefore, as
demonstrated in Ref. \cite{46}, the operators of BS1 and BS2 can be
represented using the SU(2) group theory separately in the Schwinger
representation as $B_{1}\left( -\pi /2\right) =e^{-i\frac{\pi }{2}J_{1}}$
and $B_{2}\left( \pi /2\right) =e^{i\frac{\pi }{2}J_{1}}$, respectively. In
mode $b$ of the MZI, the phase shift of the phase shifter is denoted as $%
\varphi $, and the phase shifter operator can be defined as $U\left( \varphi
\right) =e^{i\varphi b^{\dagger }b}$. The equivalent operator of the MZI can
be expressed as

\begin{equation}
U_{MZI}=e^{i\pi J_{1}/2}e^{i\varphi b^{\dagger }b}e^{-i\pi
J_{1}/2}=e^{i\varphi J_{0}}e^{-i\varphi J_{2}},  \label{7}
\end{equation}%
where the Casimir operator $J_{0}$ and the angular momentum operator $%
J_{i}\left( i=1,2,3\right) $ satisfying the commutation relation $\left[
J_{i},J_{j}\right] =i\varepsilon _{ijk}J_{k}\left( i,j,k=1,2,3\right) $ can
be represented by the photon annihilation (creation) operators $a,b$ ($%
a^{\dagger },b^{\dagger }$), as follows

\begin{eqnarray}
J_{0} &=&\frac{1}{2}\left( a^{\dag }a+b^{\dag }b\right) ,\ J_{1}=\frac{1}{2}%
\left( a^{\dag }b+ab^{\dag }\right) ,  \notag \\
J_{2} &=&\frac{1}{2i}\left( a^{\dag }b-ab^{\dag }\right) ,J_{3}=\frac{1}{2}%
\left( a^{\dag }a-b^{\dag }b\right) ,  \label{8}
\end{eqnarray}%
$\left[ J_{0},J_{i}\right] =0$, and $J_{i}$ satisfies $e^{i\frac{\pi }{2}%
J_{1}}e^{i\varphi J_{3}}e^{-i\frac{\pi }{2}J_{1}}=e^{-i\varphi J_{2}}$.
Hence, the output state $\left \vert out\right \rangle $ obtained from the
input of CS mixed with PCSVS in the MZI can be represented as

\begin{equation}
\left \vert out\right \rangle =U_{MZI}\left \vert in\right \rangle
=e^{i\varphi J_{0}}e^{-i\varphi J_{2}}\left \vert \alpha \right \rangle
_{a}\otimes \left \vert \psi _{PCSVS}\right \rangle _{b},  \label{9}
\end{equation}%
where $\left \vert in\right \rangle $ is the input state, and $\left \vert
\alpha \right \rangle _{a}=\exp [-\left \vert \alpha \right \vert
^{2}/2+\alpha a^{\dagger }]\left \vert 0\right \rangle _{a}$ is a CS on mode
$a$, $\alpha =\left \vert \alpha \right \vert e^{i\theta }$.

The parity detection at the output of mode $b$ can be defined as

\begin{equation}
\Pi _{b}=\left( -1\right) ^{b^{\dag }b}=e^{i\pi b^{\dag }b}=\int \frac{%
d^{2}\gamma }{\pi }\left \vert \gamma \right \rangle _{bb}\left \langle
-\gamma \right \vert ,\   \label{10}
\end{equation}%
where $\left \vert \gamma \right \rangle _{b}$ is the CS of mode $b$. By
utilizing Eq. (\ref{9}), the average value of parity operator is given by

\begin{eqnarray}
\left \langle \Pi _{b}\right \rangle &=&\left \langle out\right \vert \Pi
_{b}\left \vert out\right \rangle  \notag \\
&=&\left \langle in\right \vert e^{i\varphi J_{2}}e^{-i\varphi J_{0}}\Pi
_{b}e^{i\varphi J_{0}}e^{-i\varphi J_{2}}\left \vert in\right \rangle .
\label{11}
\end{eqnarray}%
As $\left[ \Pi _{b},J_{0}\right] =0$, Eq. (\ref{11}) can be reformulated as

\begin{equation}
\left \langle \Pi _{b}\right \rangle =\left \langle in\right \vert
e^{i\varphi J_{2}}\Pi _{b}e^{-i\varphi J_{2}}\left \vert in\right \rangle .\
\label{12}
\end{equation}%
Using the following unitary transformation

\begin{eqnarray}
e^{-i\varphi J_{2}}a^{\dagger }e^{i\varphi J_{2}} &=&a^{\dagger }\cos \frac{%
\varphi }{2}+b^{\dagger }\sin \frac{\varphi }{2},\   \notag \\
\ e^{-i\varphi J_{2}}b^{\dagger }e^{i\varphi J_{2}} &=&b^{\dagger }\cos
\frac{\varphi }{2}-a^{\dagger }\sin \frac{\varphi }{2},  \label{13}
\end{eqnarray}%
and $e^{-i\varphi J_{2}}\left \vert 00\right \rangle =\left \vert
00\right
\rangle $, one has

\begin{eqnarray}
e^{-i\varphi J_{2}}\left \vert in\right \rangle &=&\hat{D}^{\prime }\exp %
\left[ -\frac{1}{2}\left \vert \alpha \right \vert ^{2}+\alpha \left(
a^{\dagger }\cos \frac{\varphi }{2}+b^{\dagger }\sin \frac{\varphi }{2}%
\right) \right.  \notag \\
&&\left. +W\left( b^{\dagger }\cos \frac{\varphi }{2}-a^{\dagger }\sin \frac{%
\varphi }{2}\right) ^{2}\right] \left \vert 00\right \rangle ,  \label{14}
\end{eqnarray}%
where $\hat{D}^{\prime }$ is defined as

\begin{equation}
\hat{D}^{\prime }=\frac{W_{0}}{\sqrt{P_{m}}}\frac{\partial ^{m}}{\partial
\tau ^{m}}\frac{1}{1-\tau }\left \{ \cdot \right \} _{\tau =0}.\   \label{15}
\end{equation}%
Using Eqs. (\ref{10}) and (\ref{14}), along with the following integral
equation \cite{47}

\begin{equation}
\int \frac{d^{2}z}{\pi }e^{\zeta \left \vert z\right \vert ^{2}+\xi z+\eta
z^{\ast }+fz^{2}+gz^{\ast 2}}=\frac{e^{\frac{-\zeta \xi \eta +\xi ^{2}g+\eta
^{2}f}{\zeta ^{2}-4fg}}}{\sqrt{\zeta ^{2}-4fg}},  \label{16}
\end{equation}%
whose convergence conditions Re$\left( \xi \pm f\pm g\right) <0$ and Re$%
\left( \frac{\zeta ^{2}-4fg}{\xi \pm f\pm g}\right) <0$, Eq. (\ref{12}) can
be further calculated as

\begin{eqnarray}
\left \langle \Pi _{b}\right \rangle &=&\hat{D}\frac{\varepsilon }{\sqrt{%
1-4W_{1}W\cos ^{2}\varphi }}  \notag \\
&&\times \exp \left[ \frac{\left(
\begin{array}{c}
\left( 1-4W_{1}W\right) \left( \cos \varphi -1\right) \\
-4W_{1}W\sin ^{2}\varphi%
\end{array}%
\right) \left \vert \alpha \right \vert ^{2}}{1-4W_{1}W\cos ^{2}\varphi }%
\right]  \notag \\
&&\times \exp \left[ \frac{\left( W_{1}\alpha ^{2}+W\alpha ^{\ast 2}\right)
\sin ^{2}\varphi }{1-4W_{1}W\cos ^{2}\varphi }\right] .  \label{17}
\end{eqnarray}

In particular, for the case of $\eta =1$\ that the input state is CS mixed
with SVS \cite{33}, the above equation can be simplified as

\begin{eqnarray}
\left \langle \Pi _{b}\right \rangle &=&\frac{1}{\sqrt{1+\sinh ^{2}r\sin
^{2}\varphi }}  \notag \\
&&\times \exp \left[ \frac{2\left( \cos \varphi -1-\sinh ^{2}r\sin
^{2}\varphi \right) \left \vert \alpha \right \vert ^{2}}{2\left( 1+\sinh
^{2}r\sin ^{2}\varphi \right) }\right.  \notag \\
&&\left. -\frac{\sinh 2r\sin ^{2}\varphi \text{Re}\left( \alpha ^{2}\right)
}{2\left( 1+\sinh ^{2}r\sin ^{2}\varphi \right) }\right] .  \label{18}
\end{eqnarray}%
Since the phase $\theta =0$ of the CS can achieve better phase sensitivity
\cite{33}, here we assumes $\alpha =\alpha ^{\ast }=\left \vert \alpha
\right \vert $.

Phase sensitivity is a crucial parameter for phase measurement. Improving
phase sensitivity is synonymous with reducing phase uncertainty, leading to
increased precision in phase measurement. The phase sensitivity $\Delta
\varphi $\ with parity detection can be obtained using the average value of
the parity operator (Eq. (\ref{17})) and the error propagation formula, as
shown below:

\begin{equation}
\triangle \varphi =\frac{\triangle \Pi _{b}}{\left \vert \partial \left
\langle \Pi _{b}\right \rangle /\partial \varphi \right \vert },\
\label{19}
\end{equation}%
where $\Delta \Pi _{b}=\sqrt{\langle \Pi _{b}^{2}\rangle -\langle \Pi
_{b}\rangle ^{2}}=\sqrt{1-\langle \Pi _{b}\rangle ^{2}}$.

\begin{figure}[tbh]
\label{Fig3} \centering
\subfigure{
\begin{minipage}[b]{0.5\textwidth}
\includegraphics[width=0.83\textwidth]{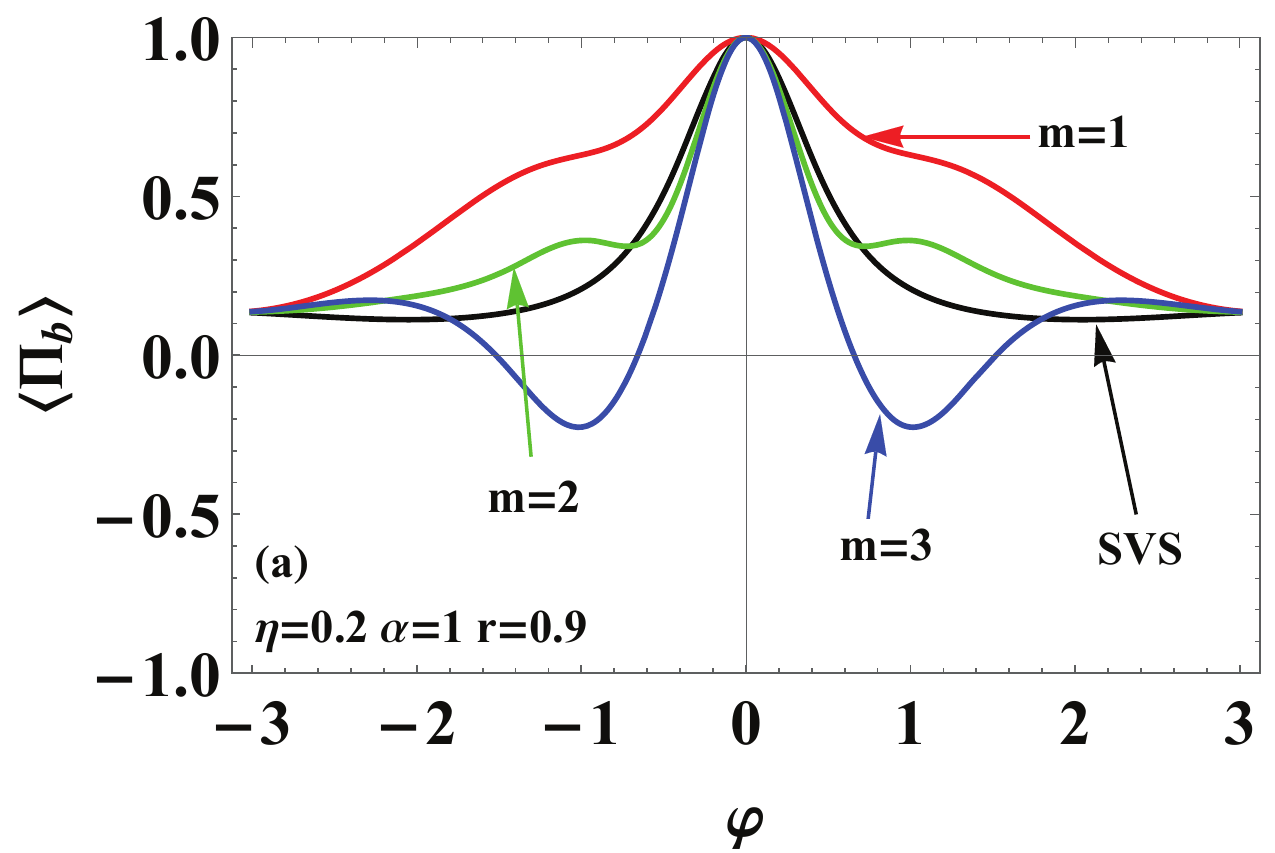}\\
\includegraphics[width=0.83\textwidth]{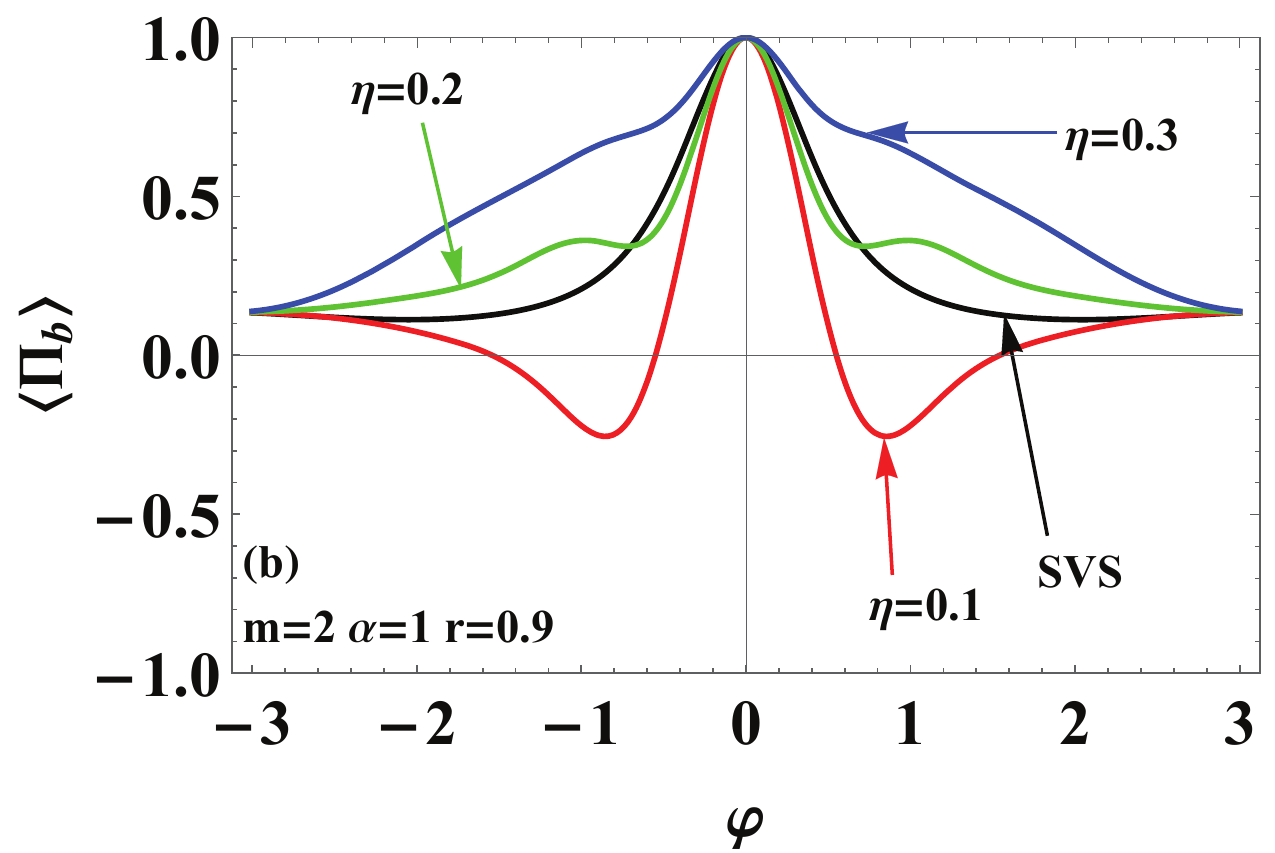}
\end{minipage}}
\caption{For a fixed coherent amplitude $\protect \alpha =1$ and squeezing
parameter $r=0.9$, (a) with transmissitivity $\protect \eta =0.2$ and
catalysis photon numbers $m=1,2,3$, and (b) with $m=2$ and $\protect \eta %
=0.1,0.2,0.3$, the variation of the parity signal $\langle \Pi _{b}\rangle $
with phase shift $\protect \varphi $. The black solid line corresponds to the
SVS (without photon catalysis) for comparison.}
\end{figure}

As depicted in Fig. 3, for a given coherent amplitude $\alpha $ and
squeezing parameter $r$, the figure illustrates the variation of the parity
signal (average value of the parity operator) with the phase shift $\varphi $%
. A narrower central peak in the image indicates a higher resolution. In
Fig. 3(a), it can be observed that, at a fixed transmissivity $\eta =0.2$ of
the BS0, the operation of multiphoton catalysis ($m\geqslant 2$) effectively
enhances the resolution of the phase shift compared to that of SVS (without
photon catalysis). Meanwhile, Fig. 3(b) illustrates that, in the case of
catalysis photon number $m=2$, adjusting $\eta $ can also result in a
narrower central peak, thus achieving better phase shift resolution than the
input CS mixed with SVS. Therefore, optimizing the parameter $\eta $ is
valid for improving the phase sensitivity.

\begin{figure}[tbh]
\label{Fig4} \centering%
\subfigure{
\begin{minipage}[b]{0.5\textwidth}
\includegraphics[width=0.83\textwidth]{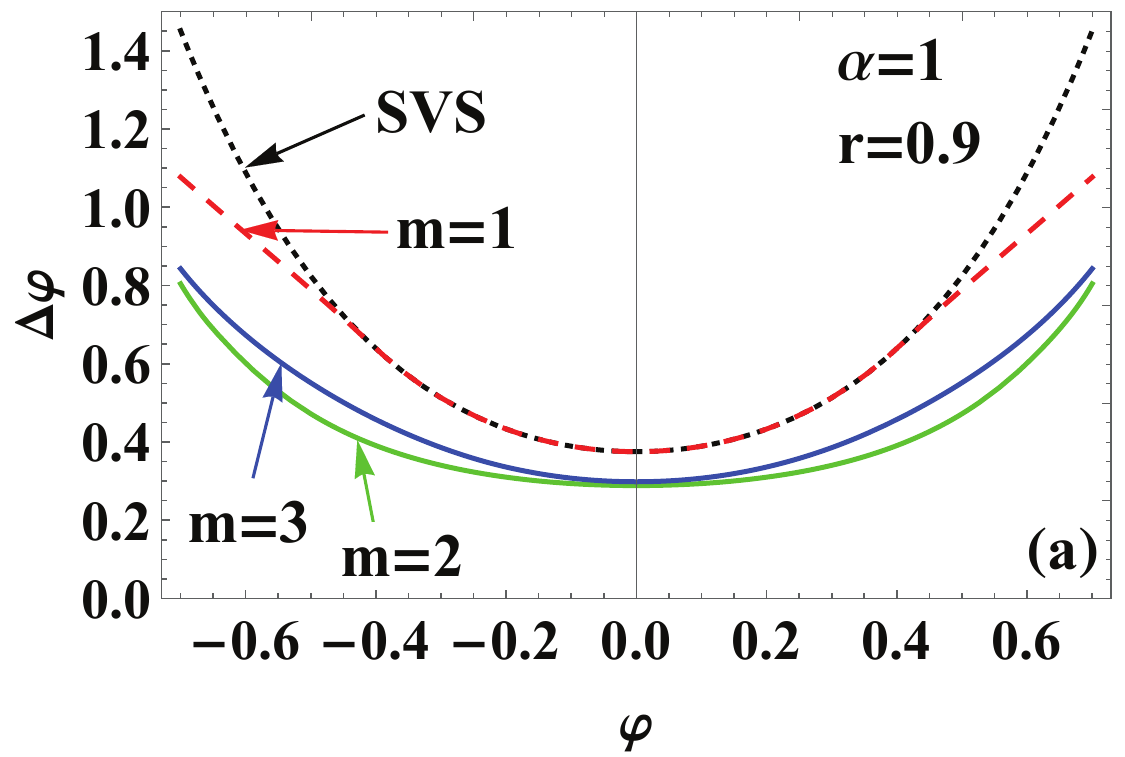}\\
\includegraphics[width=0.83\textwidth]{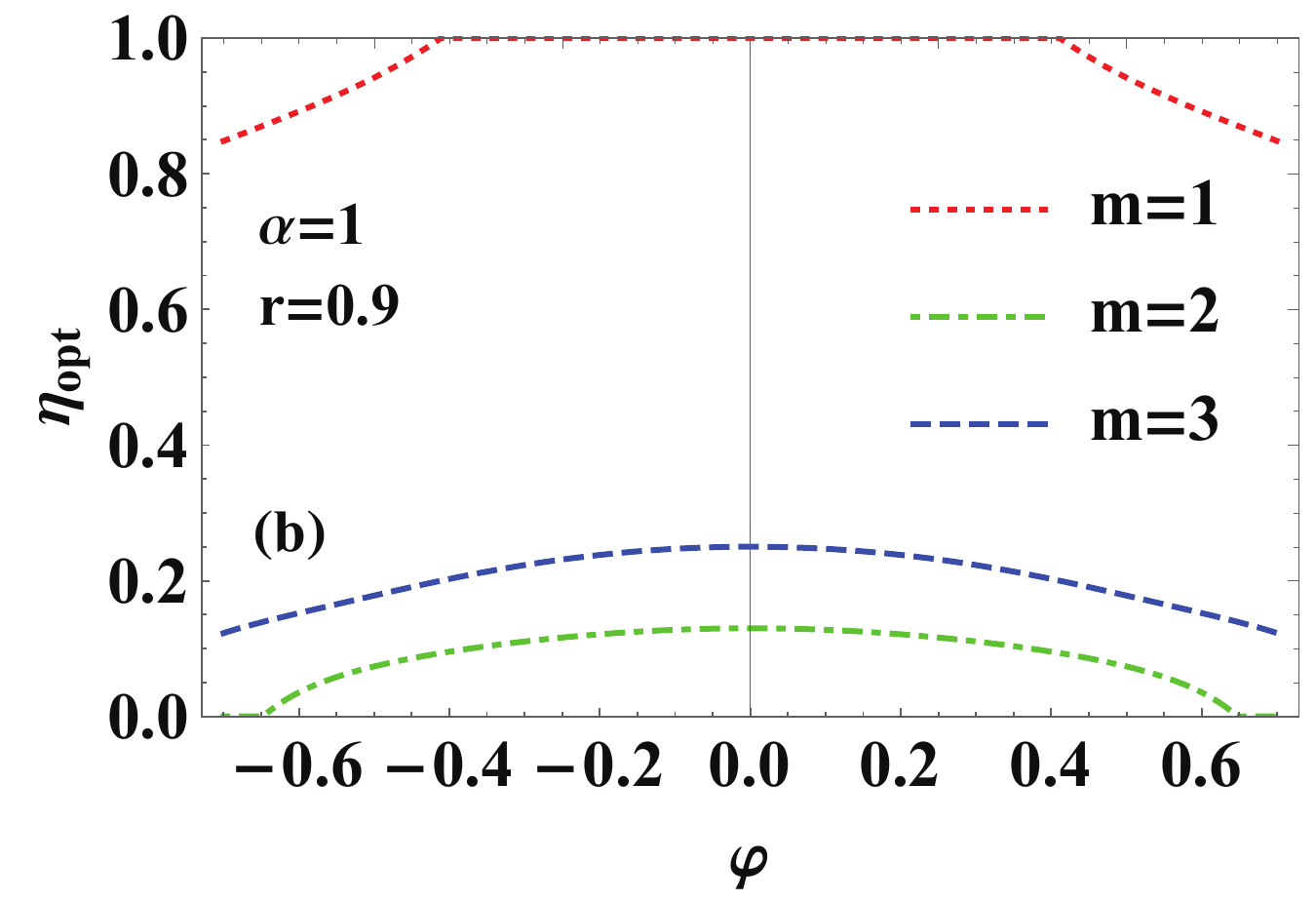}
\end{minipage}}
\caption{For the catalysis photon numbers $m=1,2,3$, a coherent amplitude $%
\protect \alpha =1$, and a squeezing parameter $r=0.9$, (a) the variation of
phase sensitivity $\Delta \protect \varphi $ with phase shift $\protect%
\varphi $ for optimized $\protect \eta $ ($\protect \eta _{opt}$), with the
black dotted line representing the SVS (without photon catalysis) for
comparision, and (b) the variation of $\protect \eta _{opt}$ with $\protect%
\varphi $.}
\end{figure}

Fig. 4(a) visually illustrates the variation of phase sensitivity $\Delta
\varphi $ with phase shift $\varphi $ for a given coherent amplitude $\alpha
=1$, squeezing parameter $r=0.9$, and optimized transmissivity $\eta _{opt}$%
. It is evident that $\Delta \varphi $ reaches its optimum value at $\varphi
=0$. For small $\varphi $ absolute values, the phase sensitivity for the
case of catalysis photon number $m=1$ is comparable to that of the CS mixed
with SVS input in the MZI. However, as the absolute value of $\varphi $
increases, the phase sensitivity for $m=1$ significantly improves.
Furthermore, for the cases of the CS mixed with the MC-SVS with $m=2$ and $3$
as input, the phase sensitivity is notably enhanced. This demonstrates the
advantageous impact of photon catalysis operation on improving phase
sensitivity with parity detection. Notably, the multiphoton catalysis
operation can notably enhance phase sensitivity. To elucidate the value of $%
\eta _{opt}$, Fig. 4(b) presents the function image of $\eta _{opt}$
changing with $\varphi $. It is apparent that for the case of $m=1$, the
optimized transmissivity $\eta _{opt}=1$, equivalent to the input state of
mode $b$\ being SVS, occurs the value of $\varphi $ between approximately $%
-0.48$ and $0.48$. Conversely, for $m=2$ and $3$, $\eta _{opt}$ is
relatively small.

\begin{figure*}[tbh]
\label{Fig5} \centering%
\subfigure{
\begin{minipage}[b]{0.83\textwidth}
\includegraphics[width=0.5\textwidth]{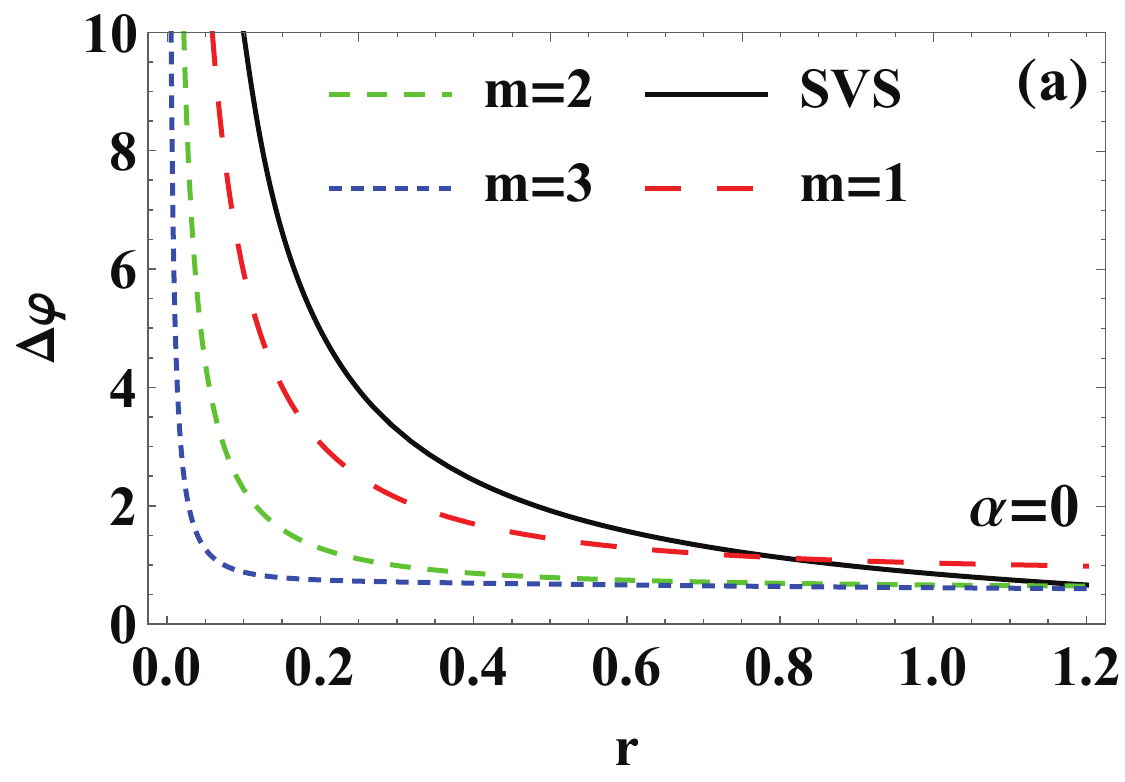}\includegraphics[width=0.5\textwidth]{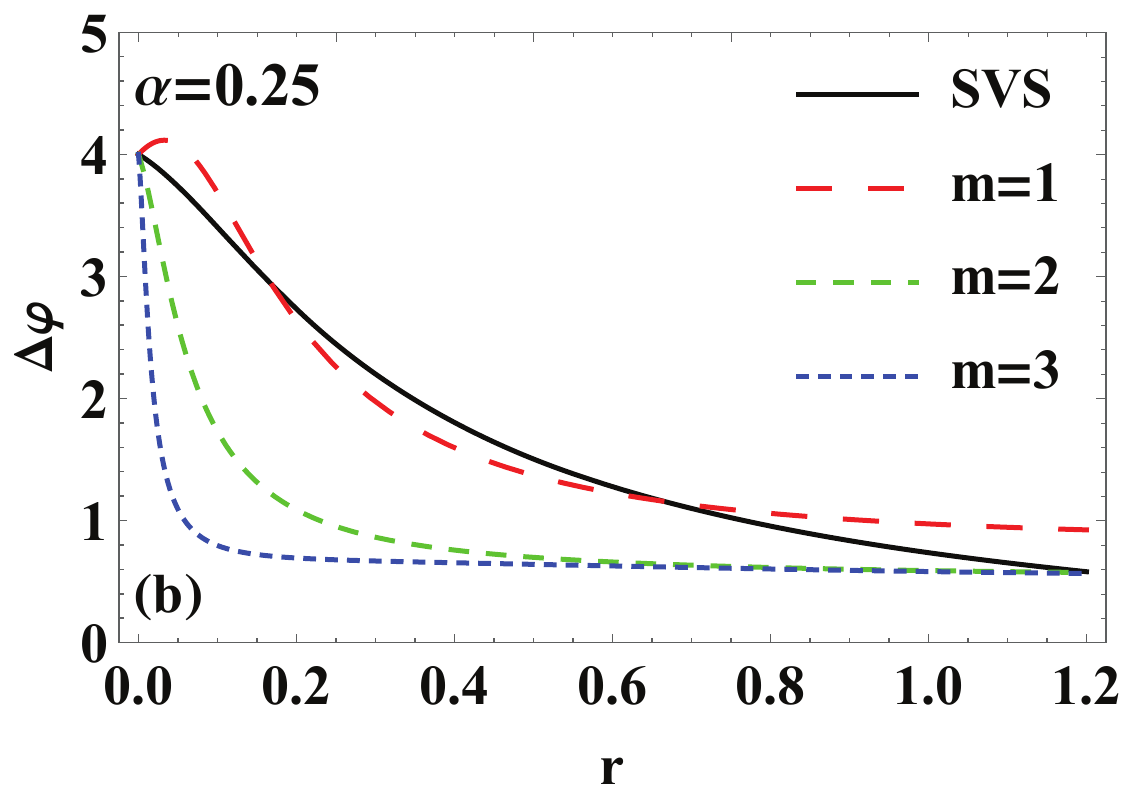}\\
\includegraphics[width=0.5\textwidth]{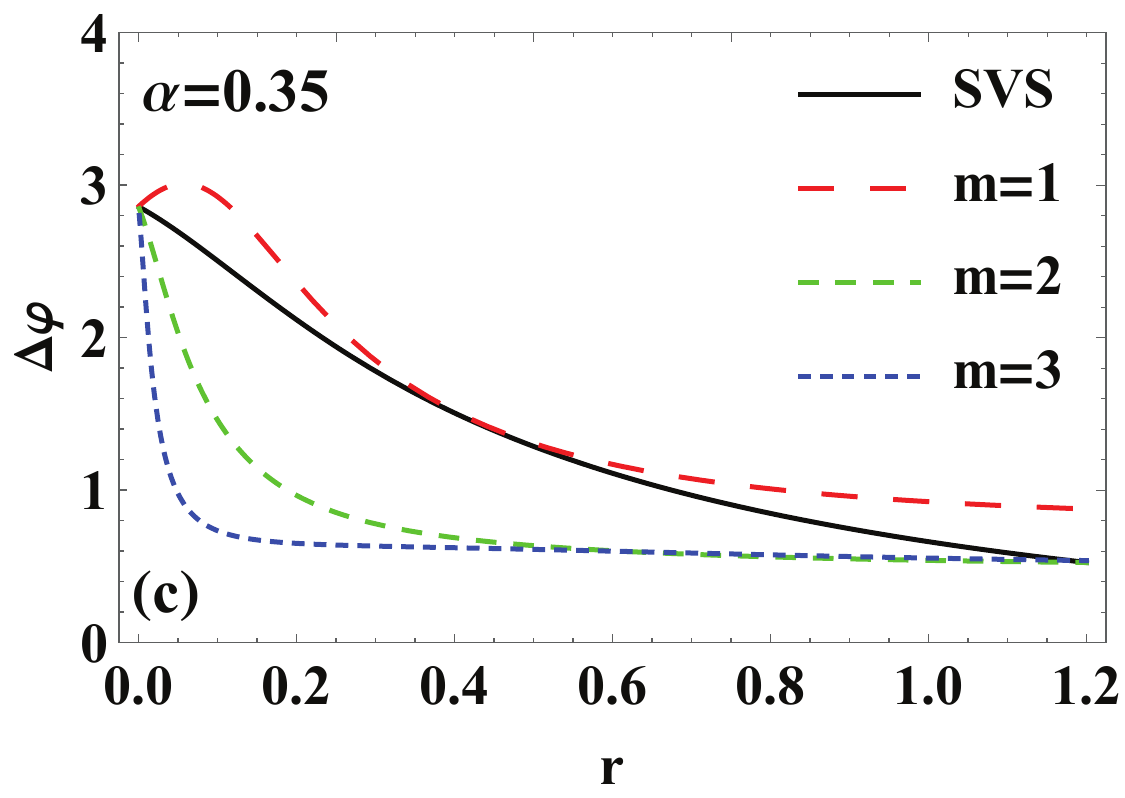}\includegraphics[width=0.5\textwidth]{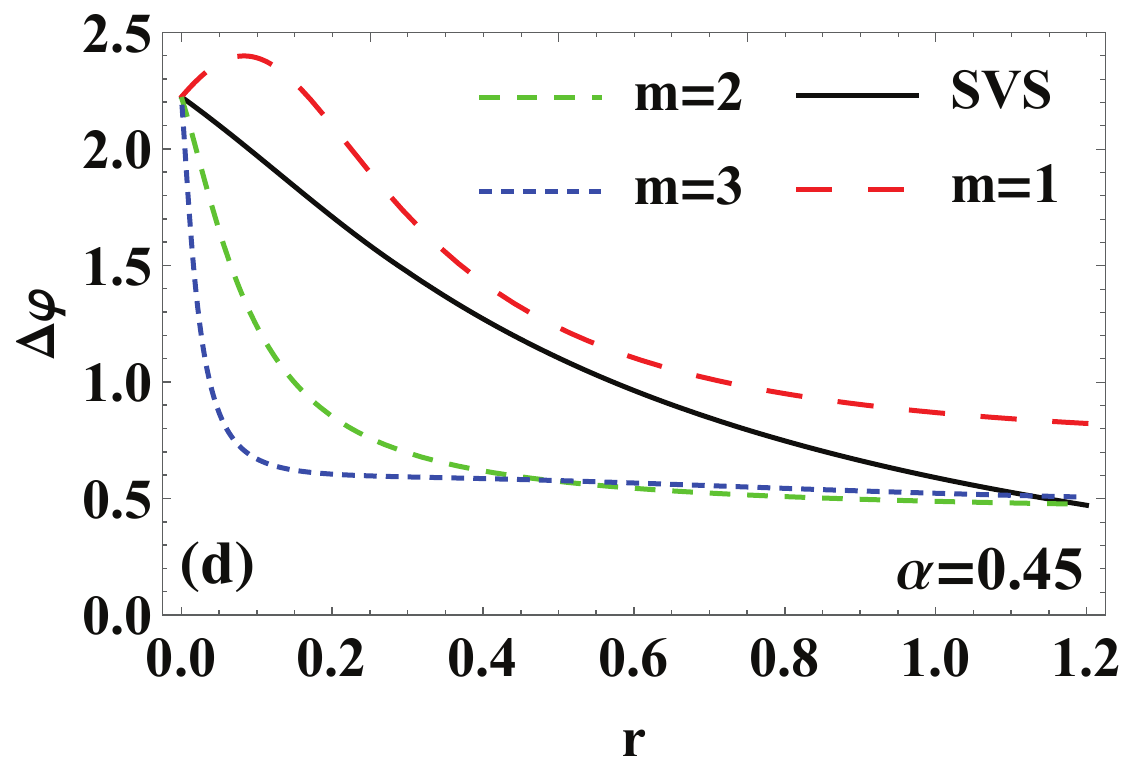}
\end{minipage}}
\caption{The phase sensitivity $\Delta \protect \varphi $ as a function of
the squeezing parameter $r$ for the catalysis photon numbers $m=1,2,3$,
fixed the phase shift $\protect \varphi =10^{-4}$, transmissitivity $\protect%
\eta =0.1$ and fixed the different coherent amplitude (a) $\protect \alpha =0$%
, (b) $\protect \alpha =0.25$, (c) $\protect \alpha =0.35$, (d) $\protect%
\alpha =0.45$. The black solid line corresponds to the SVS (without photon
catalysis) for comparison.}
\end{figure*}

To clearly see the effects of parameters of the input state on phase
sensitivity and compared with the CS and SVS as inputs, we plot the phase
sensitivity $\Delta \varphi $ as a function of the squeezing parameter $r$
for a given phase shift $\varphi =10^{-4}$, transmissivity $\eta =0.1$ and
fixed the different coherent amplitude $\alpha $, as depicted in Figs.
5(a)-(d). It is found that the phase sensitivity $\Delta \varphi $ can be
improved by increasing the value of $r$ across a wide range as well as
increasing $\alpha $. Furthermore, as can be seen in Fig. 5(a) and Fig.
5(b), when fixed the small coherent amplitude $\alpha =0,0.25$, for the case
of the catalysis photon numbers $m=1,2,3$, particularly for the case of
multiphoton catalysis operation with $m=2,3$, the phase sensitivity can be
improved more efficiently than the scheme using the CS mixed with SVS as the
input state. The relative phase sensitivity to the SVS (without photon
catalysis) can be improved within a specific range of lower squeezing
parameters for $m=1$. However, it can be seen from Fig. 5(c) and Fig. 5(d)
that increasing $\alpha $ to $0.35$ and $0.45$, respectively, leads to a
decrease in phase sensitivity for $m=1$ compared to the SVS. Nevertheless,
utilizing a CS mixed with the MC-SVS ($m=2,3$) as inputs still yields a
significant enhancement in phase sensitivity compared to using the CS mixed
with SVS. The preceding discussion leads to the conclusion that selecting
smaller values for $\alpha $, $r$, and $\eta $ would enhance the suitability
of mixing the CS with PCSVS input MZI, thereby improving phase measurement
accuracy compared to input the CS mixed with SVS. Moreover, achieving a
smaller coherent amplitude $\alpha $ and squeezing parameter $r$ is
relatively more feasible experimentally.

\begin{figure}[tbh]
\label{Fig6} \centering%
\subfigure{
\begin{minipage}[b]{0.5\textwidth}
\includegraphics[width=0.83\textwidth]{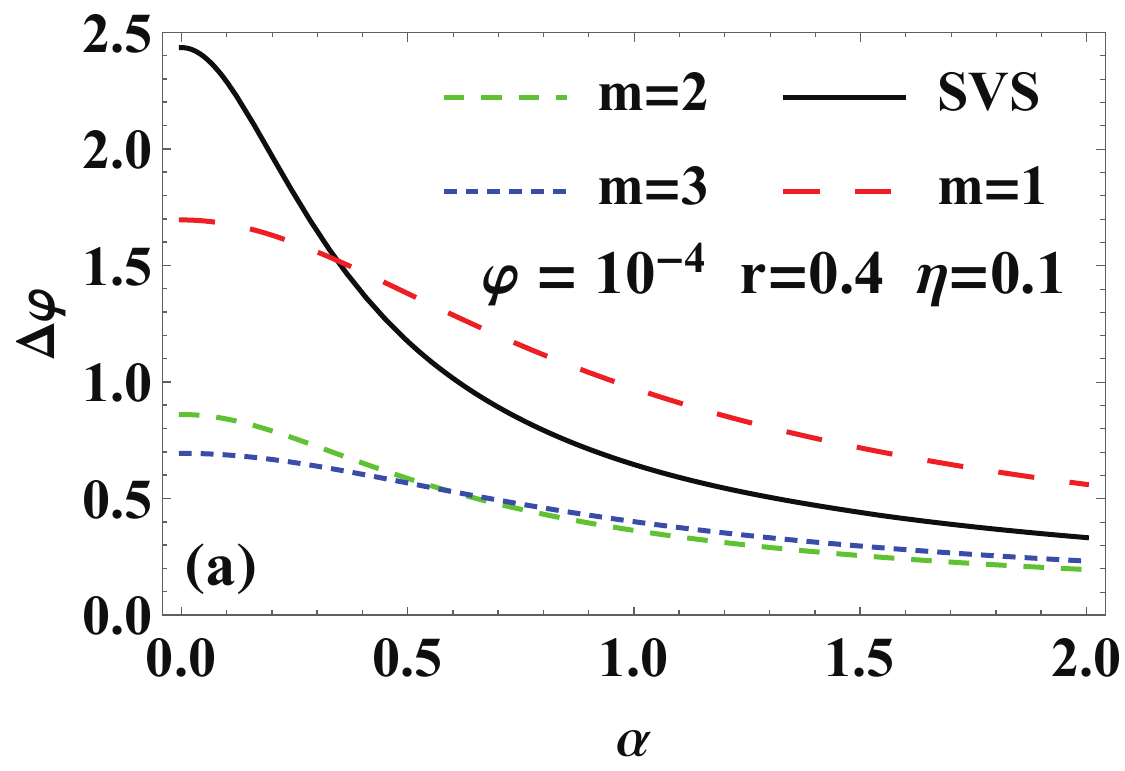}\\
\includegraphics[width=0.83\textwidth]{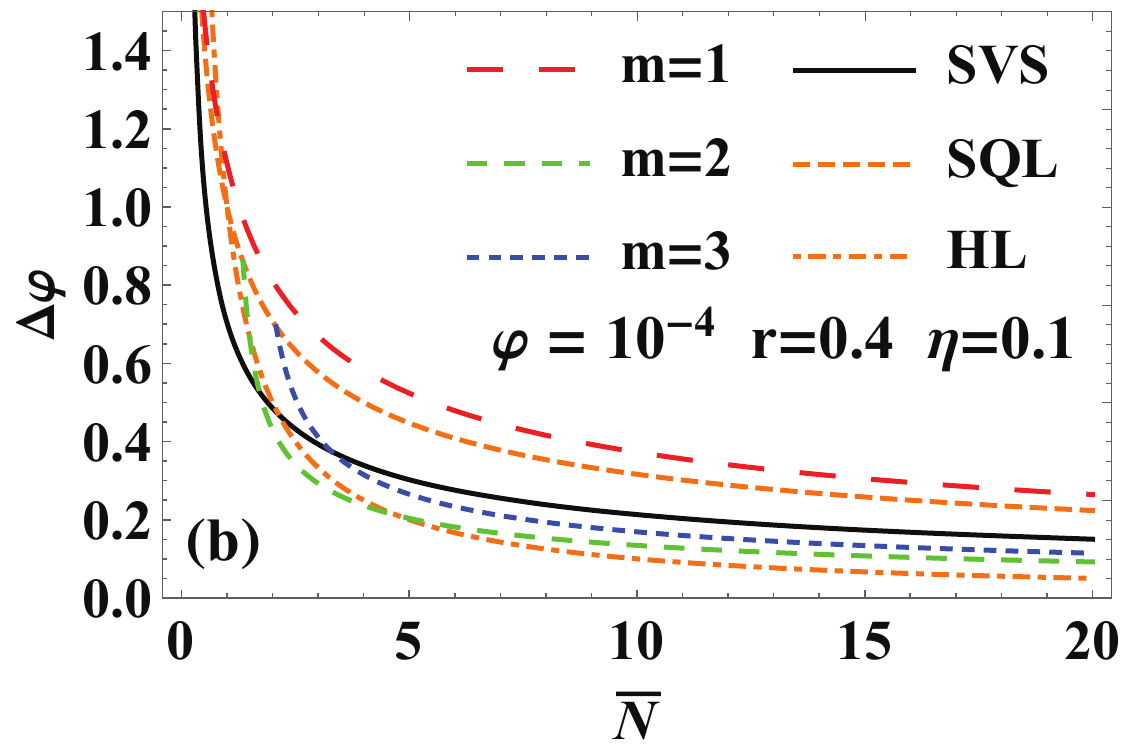}
\end{minipage}}
\caption{For the catalysis photon numbers $m=1,2,3$, fixed the phase shift $%
\protect \varphi =10^{-4}$, transmissivity $\protect \eta =0.1$ and squeezing
parameter $r=0.4$, (a) the phase sensitivity $\Delta \protect \varphi $ as a
function of the coherent amplitude $\protect \alpha $, and (b) $\Delta
\protect \varphi $ as a function of the total average photon number $\bar{N}$
comparing with the SQL and the HL. The black solid line corresponds to the
SVS (without photon catalysis) for comparison.}
\end{figure}

In Fig. 6(a), an illustration depicting the change in phase sensitivity $%
\Delta \varphi $ concerning the coherent amplitude $\alpha $ is provided for
the squeezing parameter $r=0.4$. It is apparent that $\Delta \varphi $
improves with the increase of $\alpha $, and $\Delta \varphi $ corresponding
to the catalysis photon numbers $m=1,2,3$ surpasses that of the CS mixed
with SVS as the input state. Notably, for $m=1$, $\Delta \varphi $ can be
improved relative to the case of SVS (without photon catalysis) when $\alpha
$ is small, and worse than the case of SVS when $\alpha $ is large.
Additionally, Fig. 6(b) investigates the effect of the total average photon
number $\bar{N}$ on $\Delta \varphi $. The total average photon number $\bar{%
N}$ reflects the energy of the input optical field, $\bar{N}=\bar{n}_{a}+%
\overline{n}_{b}$, where $\bar{n}_{a}=\alpha ^{2}$ represents the average
photon number of the CS input to the $a$-mode. It is evident from Fig. 6(b)
that $\Delta \varphi $ gradually enhances with the increase of $\bar{N}$,
and $\Delta \varphi $ of the multi-photon catalysis operation corresponding
to $m=2,3$ shows more significant improvement than that of the SVS.
Furthermore, in the case of $m=2,3$, $\Delta \varphi $ can surpass the SQL
and approach the HL more effectively. Specifically, for $m=2$, $\Delta
\varphi $ can exceed the HL in a small range of $\bar{N}$, and the
corresponding input resources with lower energy are easier to fabricate
experimentally.

\subsection{Quantum Fisher Information}

The QFI describes the maximum amount of information for measuring the phase
shift $\varphi $. When a pure state is input to the MZI, the QFI is
caculated as \cite{48}

\begin{equation}
F_{Q}=4\left[ \left \langle \psi _{\varphi }^{\prime }|\psi _{\varphi
}^{\prime }\right \rangle -\left \vert \left \langle \psi _{\varphi
}^{\prime }|\psi _{\varphi }\right \rangle \right \vert ^{2}\right] ,
\label{20}
\end{equation}%
where $\left \vert \psi _{\varphi }\right \rangle =e^{i\varphi b^{\dagger
}b}e^{-i\frac{\pi }{2}J_{1}}\left \vert in\right \rangle $ is the quantum
state before the BS2 of the MZI, and $\left \vert \psi _{\varphi }^{^{\prime
}}\right \rangle =\partial \left \vert \psi _{\varphi }\right \rangle
/\partial \varphi $. Further, it can be expressed using the unitary
transformations $e^{i\frac{\pi }{2}J_{1}}be^{-i\frac{\pi }{2}J_{1}}=\frac{%
\sqrt{2}}{2}\left( b-ia\right) $ and $\left \vert in\right \rangle
=\left
\vert \alpha \right \rangle _{a}\otimes \left \vert
PCSVS\right
\rangle _{b}$, and substituting Eq. (\ref{2}) into Eq. (\ref{20}%
), one can obtain

\begin{eqnarray}
F_{Q} &=&\hat{D}\epsilon \left( 1-4W_{1}W\right) ^{-\frac{3}{2}}  \notag \\
&&\times \left( 4\alpha ^{2}\left( 1-W\right) +\frac{12W_{1}W}{1-4W_{1}W}%
\right)  \notag \\
&&+\alpha ^{4}-2\alpha ^{2}-\bar{N}^{2}.\   \label{21}
\end{eqnarray}

The minimum value of phase sensitivity achievable for all measurement
schemes is known as the quantum Cram\'{e}r-Rao bound (QCRB) as defined by

\begin{equation}
\Delta \varphi _{QCRB}=\frac{1}{\sqrt{F_{Q}}}.  \label{22}
\end{equation}%
From the equation above, it can be inferred that, in theory, phase
measurement accuracy improves with an increase in QFI.

\begin{figure}[tbh]
\label{Fig7} \centering%
\subfigure{
\begin{minipage}[b]{0.5\textwidth}
\includegraphics[width=0.83\textwidth]{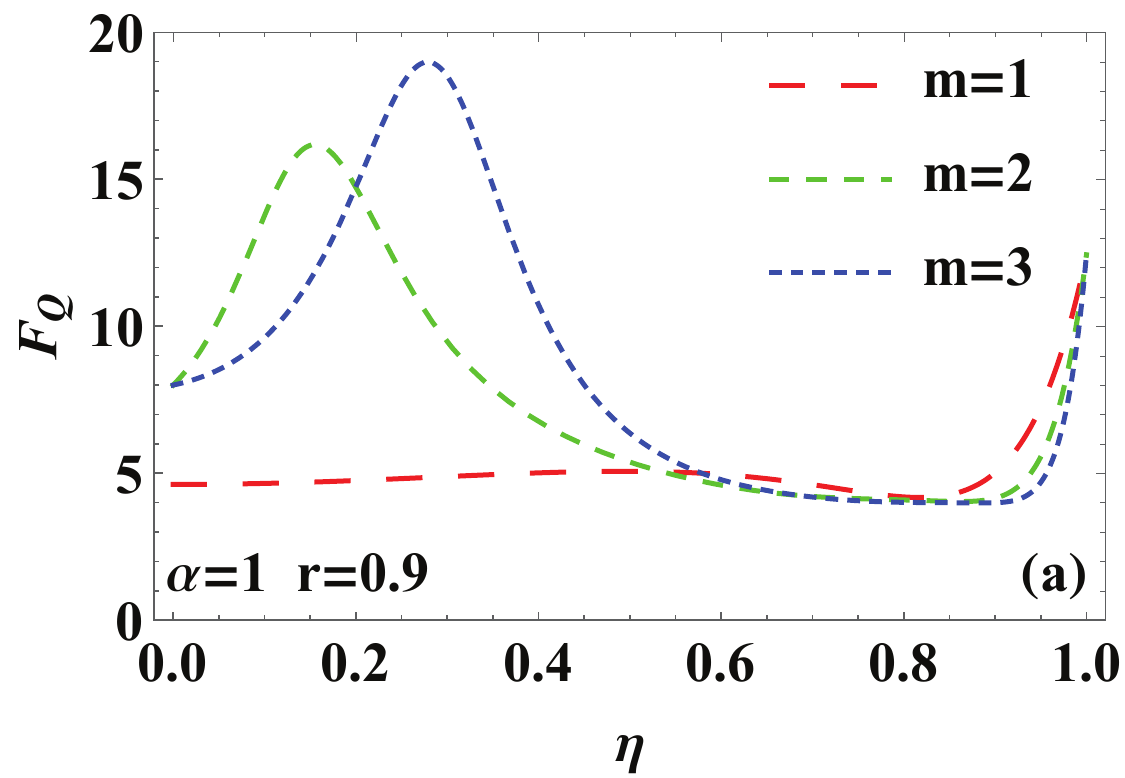}\\
\includegraphics[width=0.83\textwidth]{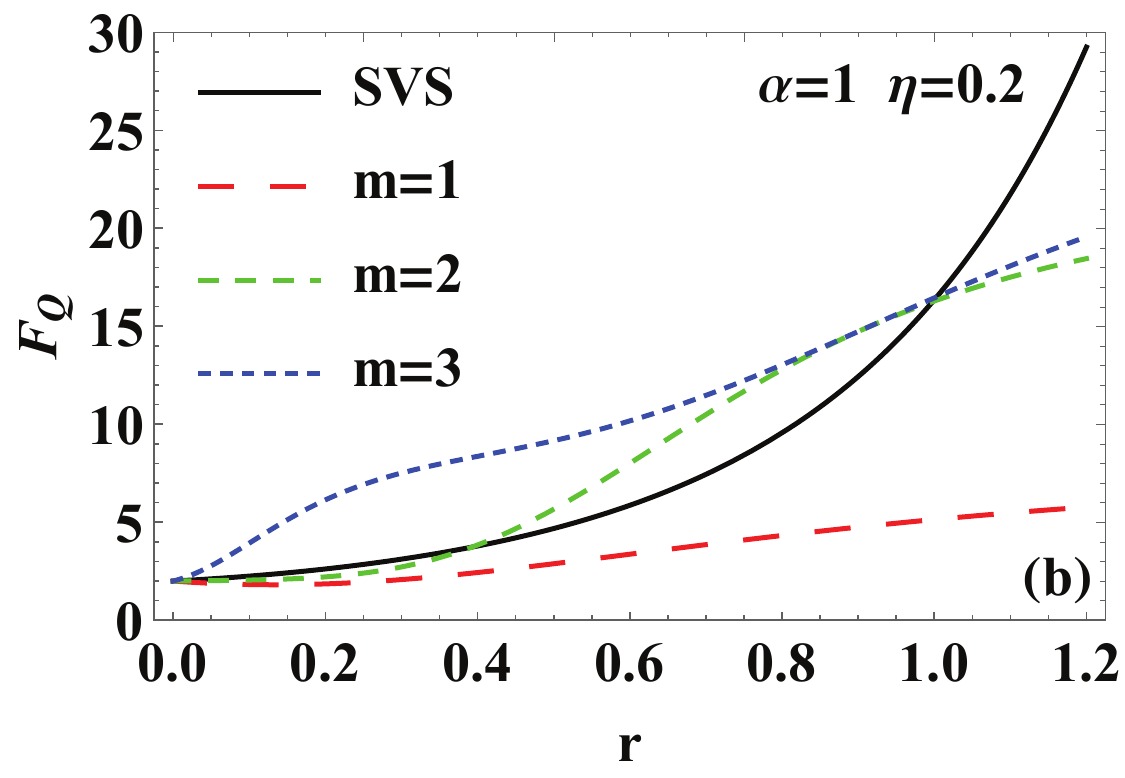}
\end{minipage}}
\caption{For the catalysis photon numbers $m=1,2,3$, (a) the variation of
QFI with the transmissivity $\protect \eta $ when fixed the coherent
amplitude $\protect \alpha =1$ and squeezing parameter $r=0.9$, (b) the
variation of QFI with the squeezing parameter $r$ when fixed $\protect \alpha %
=1$ and transmissivity $\protect \eta =0.2$, and the black solid line
corresponds to the SVS (without photon catalysis) is plotted for comparison.}
\end{figure}

Upon analysis of Fig. 7, it is evident that the QFI $F_{Q}$ varies with the
relevant parameters of the transmissivity $\eta $ and squeezing parameter $r$
of PCSVS when the catalysis photon numbers $m=1,2,3$, and is then compared
with the scenario of the CS and SVS as the input state. From Fig. 7(a), it
is evident that for a fixed squeezing parameter $r=0.9$ and coherent
amplitude $\alpha =1$, the QFI corresponding to the mixed CS and MC-SVS as
inputs ($m=2,3$) at low transmissivity $\eta $ experiences a significant
increase compared to the QFI without performing photon catalysis operation ($%
\eta =1$). This observation implies that at small $\eta $ values,
multi-photon catalysis operation can effectively enhance the precision of
phase measurement. Examining Fig. 7(b), it becomes evident that the QFI
rises as both $r$ and $m$ increase for a constant transmissivity value of $%
\eta =0.2$ and coherent amplitude $\alpha =1$. Furthermore, it is noted that
within a certain range of $r$, the QFI corresponding to the CS mixed with
the MC-SVS as the input state surpasses that of the CS mixed with SVS.

\begin{figure}[tbh]
\label{Fig8} \centering \includegraphics[width=0.83\columnwidth]{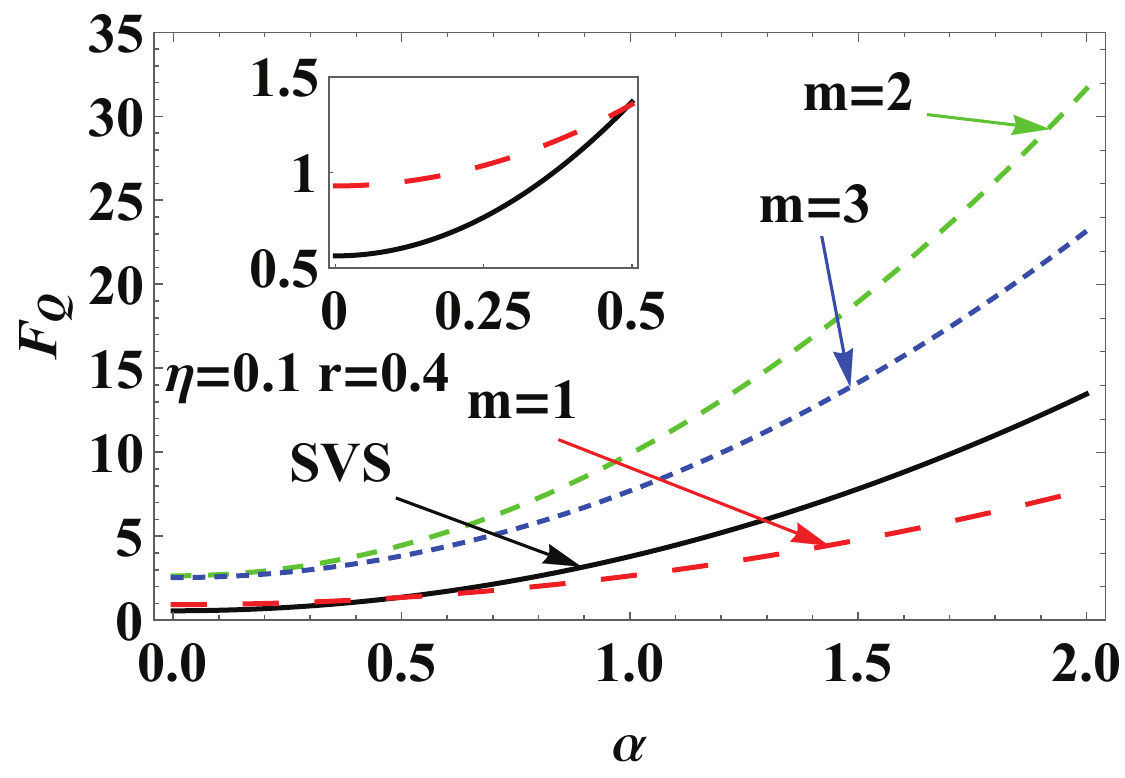}
\caption{The QFI $F_{Q}$ as a function of the coherent amplitude $\protect%
\alpha $ for the catalysis photon numbers $m=1,2,3$, transmissitivity $%
\protect \eta =0.1$ and squeezing parameter $r=0.4$. The black solid line
corresponds to the SVS (without photon catalysis) for comparison.}
\end{figure}
\  \

The plot in Fig. 8 illustrates the variation of the QFI $F_{Q}$ with the
coherent amplitude $\alpha $, given a squeezing parameter of $r=0.4$ and
transmissivity of $\eta =0.1$. It is evident from the figure that the QFI
increases as $\alpha $ increases. Furthermore, we observe that the
combination of the coherent state and the MC-SVS, with catalysis photon
numbers $m=2,3$, presents a noticeable advantage in increasing the QFI
compared to the CS mixed with SVS input MZI. Meanwhile, for the case of $m=1$%
, there is also a slight improvement when $\alpha $ is small.

\begin{figure}[tbh]
\label{Fig9} \centering%
\subfigure{
\begin{minipage}[b]{0.5\textwidth}
\includegraphics[width=0.83\textwidth]{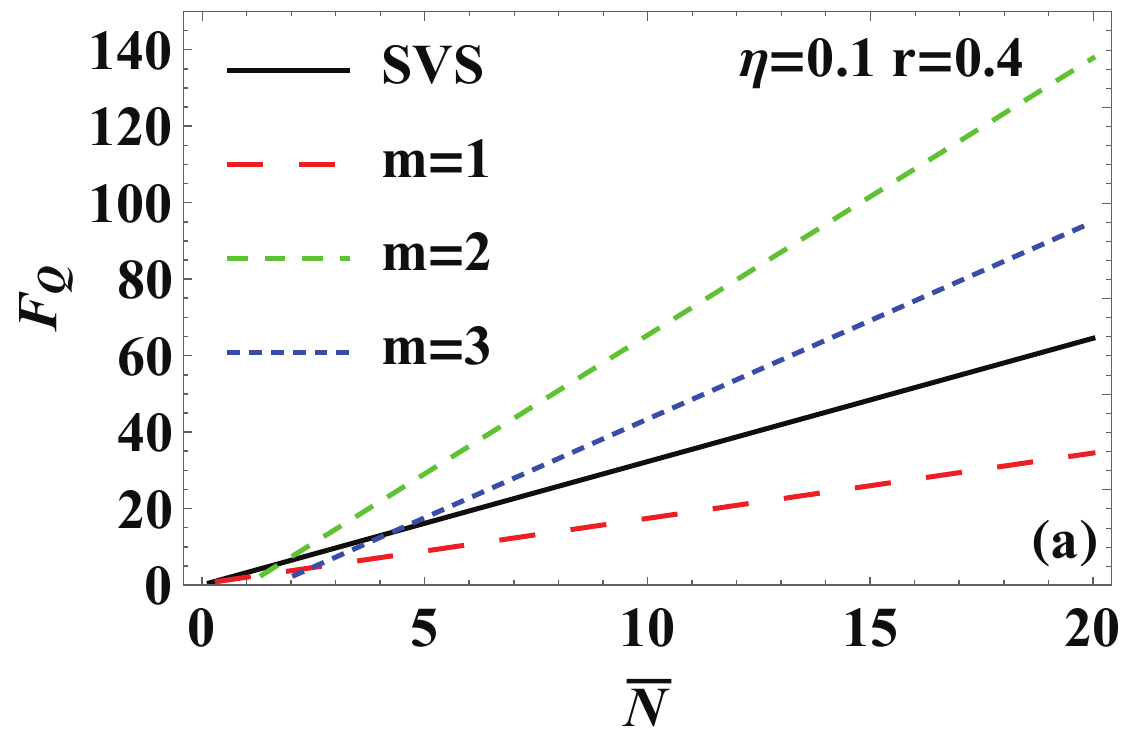}\\
\includegraphics[width=0.83\textwidth]{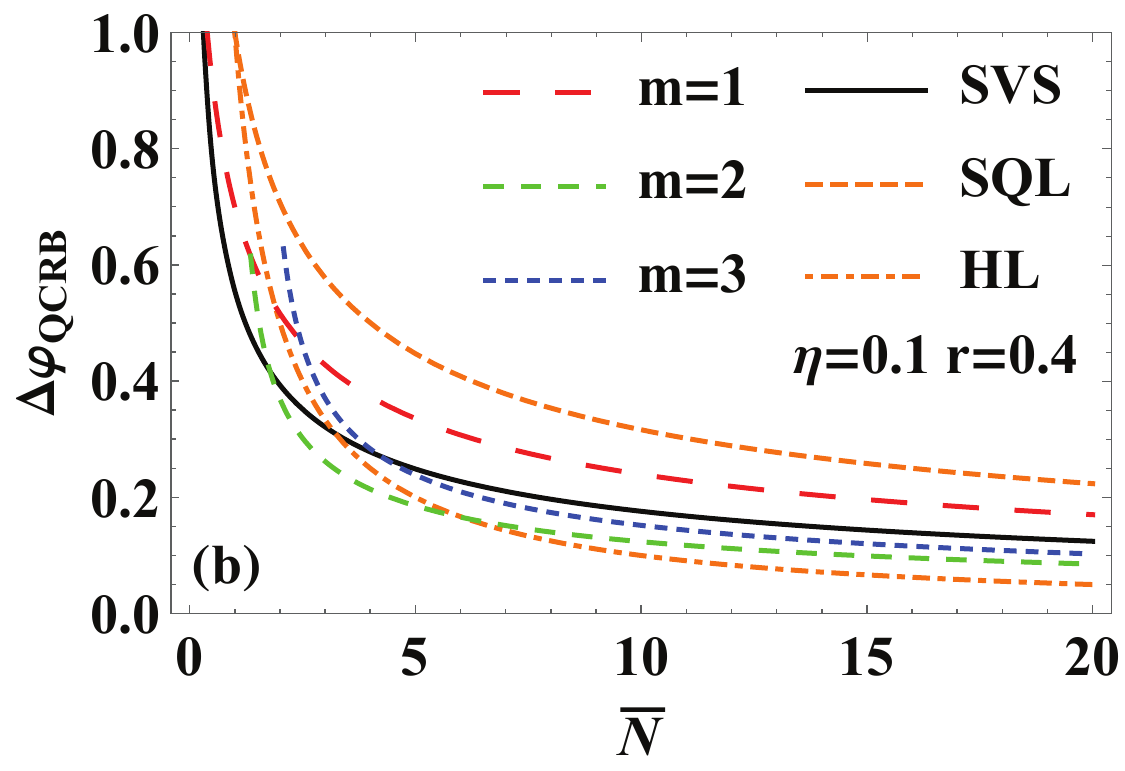}
\end{minipage}}
\caption{For the catalysis photon number $m=1,2,3$, transmissivity $\protect%
\eta =0.1$ and squeezing parameter $r=0.4$, (a) the variation of QFI with
the total average photon number $\bar{N}$, (b) the variation of QCRB with $%
\bar{N}$, the SQL, HL, and the case of SVS (without photon catalysis) are
also plotted for comparison.}
\end{figure}

To analyze the connection between phase measurement precision and the energy
of the input state, Fig. 9 illustrates the variations of QFI $F_{Q}$ and
QCRB ($\Delta \varphi _{QCRB}=1/\sqrt{F_{Q}}$) concerning the total average
photon number $\bar{N}$. In Fig. 9(a), the changes in QFI concerning $\bar{N}
$ for different catalysis photon numbers $m=1,2,3$ are depicted, with a
specific transmissivity $\eta =0.1$ and squeezing parameter $r=0.4$, and are
compared with the scenario of the CS and SVS as inputs. It is evident that
the QFI increases with $\bar{N}$, and the QFI for $m=2,3$ surpasses that of
SVS (without photon catalysis) over a wide range of $\bar{N}$. Additionally,
employing Eq. (22), the function plot of QCRB concerning $\bar{N}$ is
presented in Fig. 9(b), compared with the SQL and HL. Fig. 9(b) clearly
demonstrates that the QCRB gradually improves with increasing $\bar{N}$,
effectively surpassing the SQL. Furthermore, the QCRB can approach or even
exceed the HL when $\bar{N}$ is relatively small. Through further
comparison, it is observed that multi-photon catalysis operation for $m=2,3$
can enhance the QCRB in comparison to the SVS case.

\section{Effects of photon losses on phase sensitivity}

In practical measurement processes, photon losses are inevitably present.
Therefore, investigating the impact of photon losses on phase sensitivity is
a vital issue. In this section, we examine the effects of photon losses
before the parity detection (external losses, as depicted in Fig. 10(a)) and
between the phase shifter and the second beam splitter (BS) BS2 (internal
losses, as illustrated in Fig. 10(b)) on phase sensitivity.

\begin{figure}[tbh]
\label{Fig10} \centering \includegraphics[width=0.83\columnwidth]{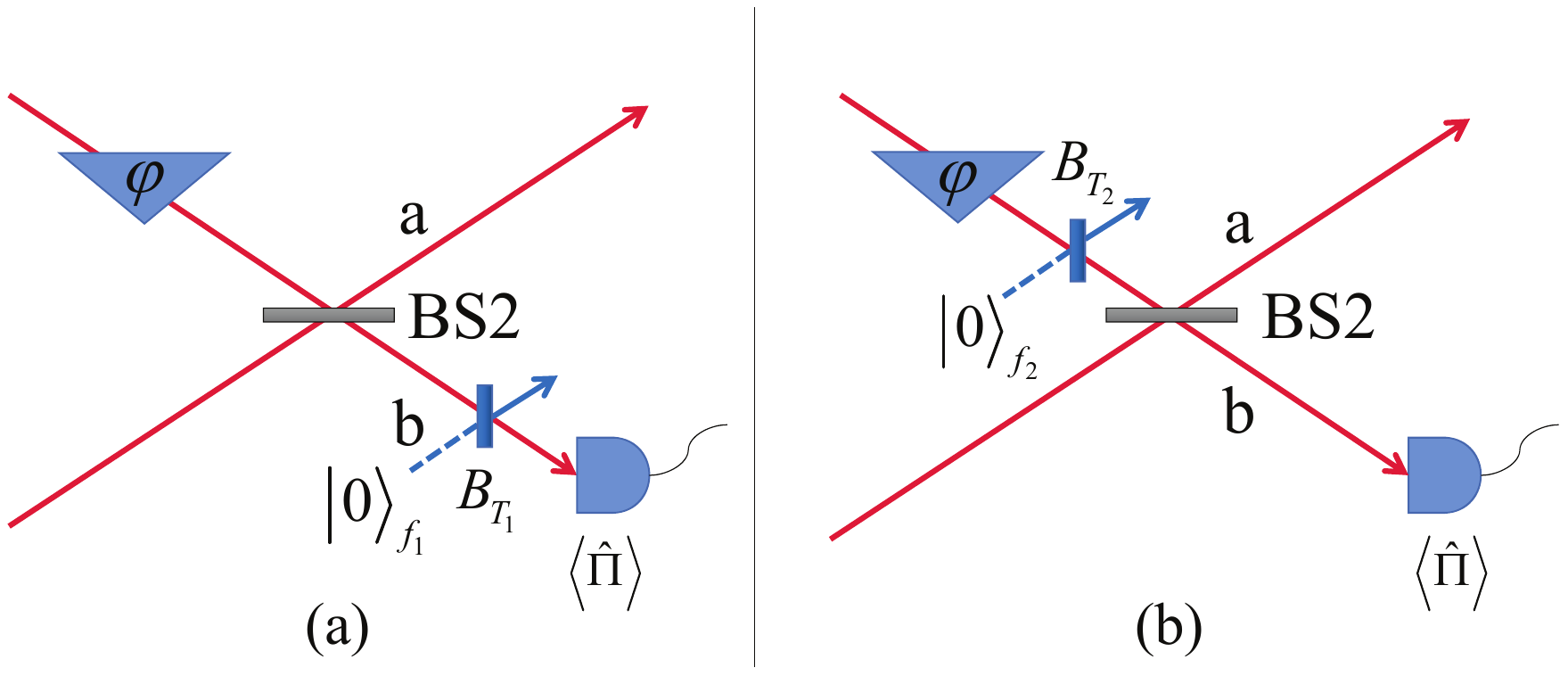}
\caption{Theoretical model of parity detection under photon losses,
including (a) external dissipation before parity detection, and (b) internal
dissipation between the phase shifter and BS2.}
\end{figure}

\subsection{Equivalent parity detection under external losses}

When considering the photon losses before parity detection (external
losses), it is essential to define the parity operator under the photon
losses in order to calculate the phase sensitivity with parity detection. As
illustrated in Fig. 10(a), the external losses can be represented using a BS
denoted as $B_{T_{1}}$. Its input-output relationship can be expressed as

\begin{equation}
B_{T_{1}}^{\dagger }\left(
\begin{array}{c}
b \\
f_{1}%
\end{array}%
\right) B_{T_{1}}=\left(
\begin{array}{cc}
\sqrt{T_{1}} & \sqrt{1-T_{1}} \\
-\sqrt{1-T_{1}} & \sqrt{T_{1}}%
\end{array}%
\right) \left(
\begin{array}{c}
b \\
f_{1}%
\end{array}%
\right) ,  \label{23}
\end{equation}%
where $b$ ($b^{\dagger }$) and $f_{1}$ ($f_{1}^{\dagger }$) are the photon
annihilation (creation) operators corresponding to the mode $b$ in MZI and
the dissipative mode $f_{1}$ respectively, and $T_{1}$ represents the
transmissivity of the BS, ranging from $0$ to $1$. The larger the value of $%
T_{1}$, the smaller the photon losses. $T_{1}=1$ corresponds to the ideal
situation without losses. To derive the phase sensitivity in the presence of
external losses, it is necessary to initially express the parity operator in
the ideal case as a Weyl ordering form \cite{49}, i.e.,

\begin{equation}
\Pi _{b}=\frac{\pi }{2}%
\begin{array}{c}
\colon \\
\colon%
\end{array}%
\delta \left( b\right) \delta \left( b^{\dag }\right)
\begin{array}{c}
\colon \\
\colon%
\end{array}%
,\   \label{24}
\end{equation}%
where$%
\begin{array}{c}
\colon \\
\colon%
\end{array}%
\bullet
\begin{array}{c}
\colon \\
\colon%
\end{array}%
$ is the Weyl ordering and $\delta \left( \cdot \right) $ represents the
delta function. Combining Eqs. (\ref{23}) and (\ref{24}), using the Weyl
invariance under similarity transformations \cite{50,51}, and considering
the vacuum noise $\left \vert 0\right \rangle _{f_{1}}$ at the input of the
dissipative mode $f_{1}$, the parity operator under external losses can be
expressed as

\begin{eqnarray}
\Pi _{b}^{loss} &=&\frac{\pi }{2}\text{Tr}\left[ \left \vert 0\right \rangle
_{f_{1}}\left \langle 0\right \vert
\begin{array}{c}
\colon \\
\colon%
\end{array}%
B_{T_{1}}^{\dagger }\delta \left( b\right) \delta \left( b^{\dag }\right)
B_{T_{1}}%
\begin{array}{c}
\colon \\
\colon%
\end{array}%
\right]  \notag \\
&=&\frac{\pi }{2}\text{Tr}\left[ \left \vert 0\right \rangle _{f_{1}}\left
\langle 0\right \vert
\begin{array}{c}
\colon \\
\colon%
\end{array}%
\delta (\sqrt{T_{1}}b+\sqrt{1-T_{1}}f_{1})\right.  \notag \\
&&\left. \times \delta \left( \sqrt{T_{1}}b^{\dag }+\sqrt{1-T_{1}}%
f_{1}^{\dag }\right)
\begin{array}{c}
\colon \\
\colon%
\end{array}%
\right] .  \label{25}
\end{eqnarray}%
Utilizing the normal product form of the Wigner operator \cite{52,53} $%
\Delta _{\alpha }\left( \alpha \right) =\frac{1}{\pi }\colon \exp \left[
-2\left( a-\alpha \right) \left( a^{\dagger }-\alpha ^{\ast }\right) \right]
\colon $and $\Delta _{\beta }\left( \beta \right) =\frac{1}{\pi }\colon \exp %
\left[ -2\left( b-\beta \right) \left( b^{\dagger }-\beta ^{\ast }\right) %
\right] \colon $, the classical correspondence of Weyl ordering operator can
be expressed as follows \cite{51}

\begin{eqnarray}
&&%
\begin{array}{c}
\colon \\
\colon%
\end{array}%
f\left( a,a^{\dag },b,b^{\dag }\right)
\begin{array}{c}
\colon \\
\colon%
\end{array}
\notag \\
&=&4\int d^{2}\alpha d^{2}\beta f\left( \alpha ,\alpha ^{\ast },\beta ,\beta
^{\ast }\right) \Delta _{a}\left( \alpha \right) \Delta _{b}\left( \beta
\right) .\   \label{26}
\end{eqnarray}%
According to Eq. (\ref{26}), $\Pi _{b}^{loss}$ can be further derived based
on Eq. (\ref{25}) as follows

\begin{eqnarray}
\Pi _{b}^{loss} &=&2\pi \int \frac{d^{2}\beta d^{2}\gamma }{\pi ^{2}}\delta (%
\sqrt{T_{1}}\beta +\sqrt{1-T_{1}}\gamma )  \notag \\
&&\times \delta (\sqrt{T_{1}}\beta ^{\ast }+\sqrt{1-T_{1}}\gamma ^{\ast })
\notag \\
&&\times Tr\left[ \left \vert 0\right \rangle _{f_{1}}\left \langle 0\right
\vert \Delta _{\alpha }\left( \alpha \right) \Delta _{\beta }\left( \beta
\right) \right]  \notag \\
&=&\colon \exp \left( -2T_{1}b^{\dagger }b\right) \colon =\left(
1-2T_{1}\right) ^{b^{\dagger }b}.\   \label{27}
\end{eqnarray}

Based on Eqs. (\ref{9}) and (\ref{27}), the average value of the parity
operator under external losses for the output state of MZI can be derived as

\begin{equation}
\left \langle \Pi _{b}^{loss}\right \rangle =\hat{D}\frac{\varepsilon \exp %
\left[ \frac{A_{1}T_{1}\left( \cos \varphi -1\right) +A_{2}}{A_{1}}\alpha
^{2}\right] }{\sqrt{A_{1}}},\   \label{28}
\end{equation}%
where

\begin{eqnarray}
A_{1} &=&1-4W_{1}W\left[ 1-T_{1}\left( \cos \varphi +1\right) \right] ^{2},
\notag \\
A_{2} &=&2WT_{1}^{2}\sin ^{2}\varphi  \notag \\
&&-4W_{1}WT_{1}^{2}\sin ^{2}\varphi \left[ T_{1}\left( \cos \varphi
+1\right) -1\right] .\   \label{29}
\end{eqnarray}%
The phase sensitivity $\Delta \varphi $ in the case of external dissipation
can be obtained using an error propagation formula similar to Eq. (\ref{19}).

\subsection{Equivalent parity detection under internal losses}

As depicted in Fig. 10(b), the process of photon losses (internal losses)
between the phase shifter and BS2 can be simulated by the BS $B_{T_{2}}$,
where the input states of $B_{T_{2}}$ correspond to the signal light of mode
$b$ and vacuum noise $\left \vert 0\right \rangle _{f_{2}}$ of dissipation
mode $f_{2}$. The corresponding input-output relation is given by

\begin{equation}
B_{T_{2}}^{\dagger }\left(
\begin{array}{c}
b \\
f_{2}%
\end{array}%
\right) B_{T_{2}}=\left(
\begin{array}{cc}
\sqrt{T_{2}} & \sqrt{1-T_{2}} \\
-\sqrt{1-T_{2}} & \sqrt{T_{2}}%
\end{array}%
\right) \left(
\begin{array}{c}
b \\
f_{2}%
\end{array}%
\right) ,  \label{30}
\end{equation}%
where $b$ ($b^{\dagger }$) and $f_{2}$ ($f_{2}^{\dagger }$) are the photon
annihilation (creation) operators corresponding to the mode $b$ and the
dissipative mode $f_{2}$, respectively. $T_{2}$ represents the
transmissivity of the beam splitter $B_{T_{2}}$, which ranges from $0$ to $1$%
. The larger the value of $T_{2}$, the smaller the photon losses. $T_{2}=1$
corresponds to the ideal situation of no losses. Therefore, in the
Heisenberg picture, for the case of internal dissipation, by using the
transformation relation of the equivalent operator of MZI considering
internal dissipation, the operator representing parity detection of the
input state after passing through the entire lossy interferometer can be
given by

\begin{equation}
\widetilde{\Pi }_{b}^{loss}=\left. _{f_{2}}\left \langle 0\right \vert
B_{1}^{\dagger }U^{\dagger }\left( \varphi \right) B_{T_{2}}^{\dagger
}B_{2}^{\dagger }\Pi _{b}B_{2}B_{T_{2}}U\left( \varphi \right) B_{1}\left
\vert 0\right \rangle _{f_{2}}\right. .  \label{31}
\end{equation}%
Applying the method similar to that used to obtain the parity operator in
the case of external dissipation, and combining Eq. (\ref{31}) with the
transformation relationship of the two $50:50$ BSs in the MZI

\begin{eqnarray}
B_{1}^{\dagger }\binom{a}{b}B_{1} &=&\frac{\sqrt{2}}{2}\left(
\begin{array}{cc}
1 & -i \\
-i & 1%
\end{array}%
\right) \binom{a}{b},  \notag \\
B_{2}^{\dagger }\binom{a}{b}B_{2} &=&\frac{\sqrt{2}}{2}\left(
\begin{array}{cc}
1 & i \\
i & 1%
\end{array}%
\right) \binom{a}{b},  \label{32}
\end{eqnarray}%
the expression for the normal-ordered parity operator in the case of
internal dissipation can be obtained as follows

\begin{equation}
\widetilde{\Pi }_{b}^{loss}=\colon e^{X_{1}a^{\dagger }a-X_{2}b^{\dagger
}a-X_{2}^{\ast }a^{\dagger }b+X_{3}b^{\dagger }b}\colon ,  \label{33}
\end{equation}%
where

\begin{eqnarray}
X_{1} &=&\frac{2\sqrt{T_{2}}\cos \varphi -1-T_{2}}{2},  \notag \\
X_{2} &=&\frac{\left( T_{2}+1\right) ^{2}-4T_{2}\cos ^{2}\varphi }{2\left(
iT_{2}-i+2\sqrt{T_{2}}\sin \varphi \right) },  \notag \\
X_{3} &=&-\frac{2\sqrt{T_{2}}\cos \varphi +1+T_{2}}{2}.  \label{34}
\end{eqnarray}

Upon combining with the expression of the input states, the average value of
the corresponding parity operator in the case of internal dissipation can be
further obtained as follows

\begin{equation}
\left \langle \widetilde{\Pi }_{b}^{loss}\right \rangle =\left \langle
in\right \vert \widetilde{\Pi }_{b}^{loss}\left \vert in\right \rangle =\hat{%
D}\frac{\varepsilon \exp \left[ \frac{B_{1}X_{1}+B_{2}}{B_{1}}\alpha ^{2}%
\right] }{\sqrt{B_{1}}},  \label{35}
\end{equation}%
where

\begin{eqnarray}
B_{1} &=&1-4W_{1}W\left( 1+X_{3}\right) ^{2},  \notag \\
B_{2} &=&W_{1}X_{2}^{2}+WX_{2}^{\ast 2}+4W_{1}W\left \vert X_{2}\right \vert
^{2}\left( 1+X_{3}\right) .  \label{36}
\end{eqnarray}%
Further substituting of Eq. (\ref{35}) into the error propagation formula,
we can obtain the phase sensitivity $\Delta \varphi _{L}$ with parity
detection in the case of internal dissipation.

\begin{figure}[tbh]
\label{Fig11} \centering%
\subfigure{
\begin{minipage}[b]{0.5\textwidth}
\includegraphics[width=0.83\textwidth]{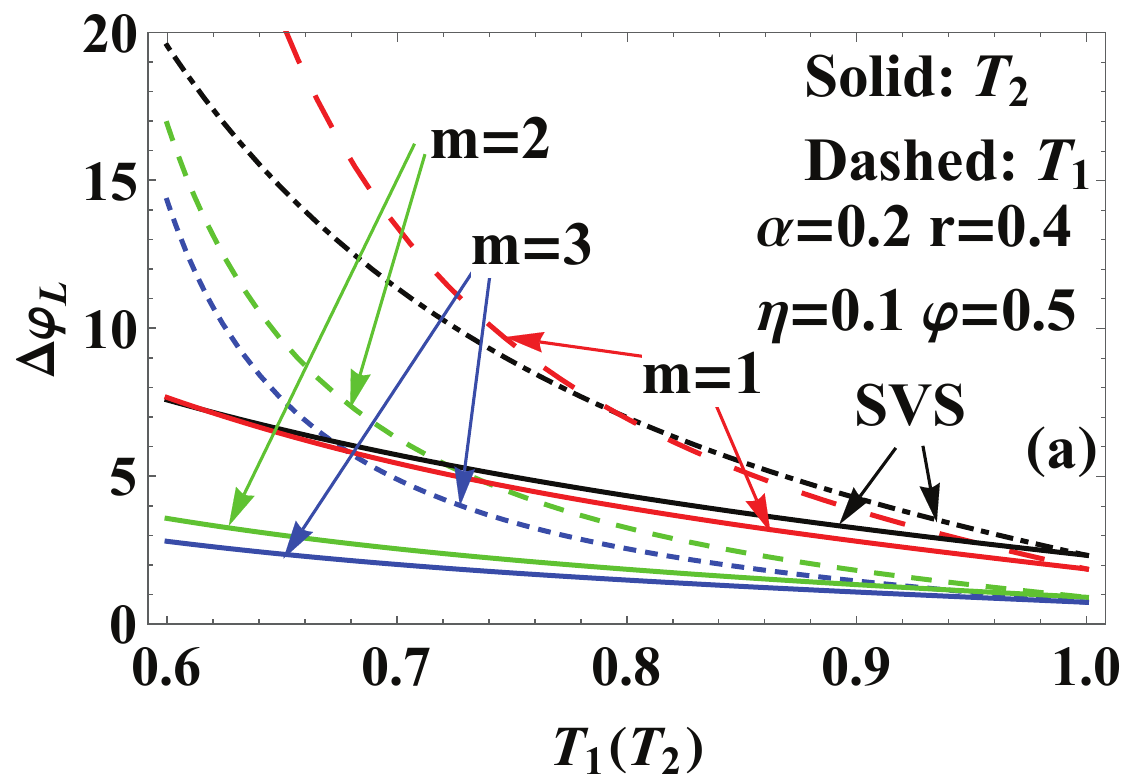}\\
\includegraphics[width=0.83\textwidth]{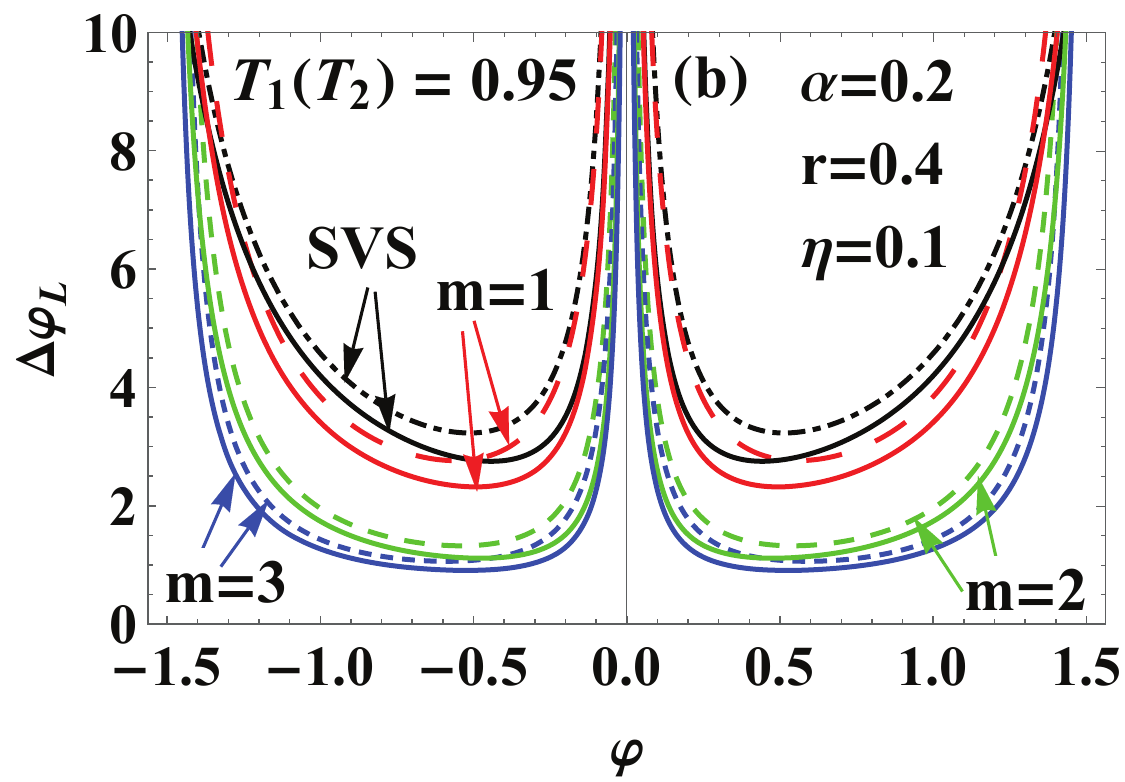}
\end{minipage}}
\caption{For the catalysis photon numbers $m=1,2,3$, and fixed the
transmissivity $\protect \eta =0.1$, the squeezing parameter $r=0.4$, and the
coherent amplitude $\protect \alpha =0.2$, (a) the phase sensitivity $\Delta
\protect \varphi _{L}$ under photon losses as a function of the
transmissivity of $B_{T_{1}}$($B_{T_{2}}$) $T_{1}$($T_{2}$) for the phase
shift $\protect \varphi =0.5$, (b) $\Delta \protect \varphi _{L}$ as a
function of $\protect \varphi $ for $T_{1}=T_{2}=0.95$. The dashed (solid)
lines correspond to the cases of external (internal) dissipation, and the
black lines correspond to the SVS (without photon catalysis) for comparison.}
\end{figure}

\subsection{The effects of external and internal losses}

In order to clearly analyze the impact of photon losses on phase
sensitivity, an image of the phase sensitivity $\Delta \varphi _{L}$ under
photon losses is presented as a function of the transmissivity $T_{1}$($%
T_{2} $) of the optical BS $B_{T_{1}}$($B_{T_{2}}$) in Fig. 11(a) for the
different catalysis photon numbers $m=1,2,3$, phase shift $\varphi =0.5$,
and fixed the other relevant parameters ($\eta ,r,\alpha $) of the input
state, and comparing it with the SVS (without photon catalysis). It is
evident that $\Delta \varphi _{L}$ increases as $T_{1}$($T_{2}$) decreases,
indicating that photon losses worsen the phase sensitivity. Fig. 11(b)
demonstrates the variation of $\Delta \varphi _{L}$ with $\varphi $ for $%
m=1,2$, and $3$, and $\eta =0.1$, when $r=0.4$, $\alpha =0.2$, and $%
T_{1}=T_{2}=0.95$. These results are compared with those of the CS mixed
with SVS as\ inputs. Fig. 11 suggests that, under photon losses, the optimal
value of $\Delta \varphi _{L}$ is achieved when $\varphi $ deviates from $0$%
; conversely, the value of $\Delta \varphi _{L}$ becomes extremely large
when $\varphi $ approaches $0$. Additionally, it is evident from Fig. 11
that the phase sensitivity of the external dissipation, corresponding to the
dashed line, is inferior to that of the internal dissipation, corresponding
to the solid line, indicating a greater impact of external dissipation on
phase sensitivity. Despite photon losses in the actual measurement process,
especially for the MC-SVS with $m=2,3$, $\Delta \varphi _{L}$ can still be
efficiently improved compared to the CS mixed with SVS when using a mix of
CS and the PCSVS as inputs. Furthermore, for external dissipation, a smaller
loss (corresponding to a larger $T_{1}$) is required to achieve better phase
sensitivity at $m=1$ than the SVS case (see Fig. 11(a)).

\begin{figure}[tbh]
\label{Fig12} \centering%
\subfigure{
\begin{minipage}[b]{0.5\textwidth}
\includegraphics[width=0.83\textwidth]{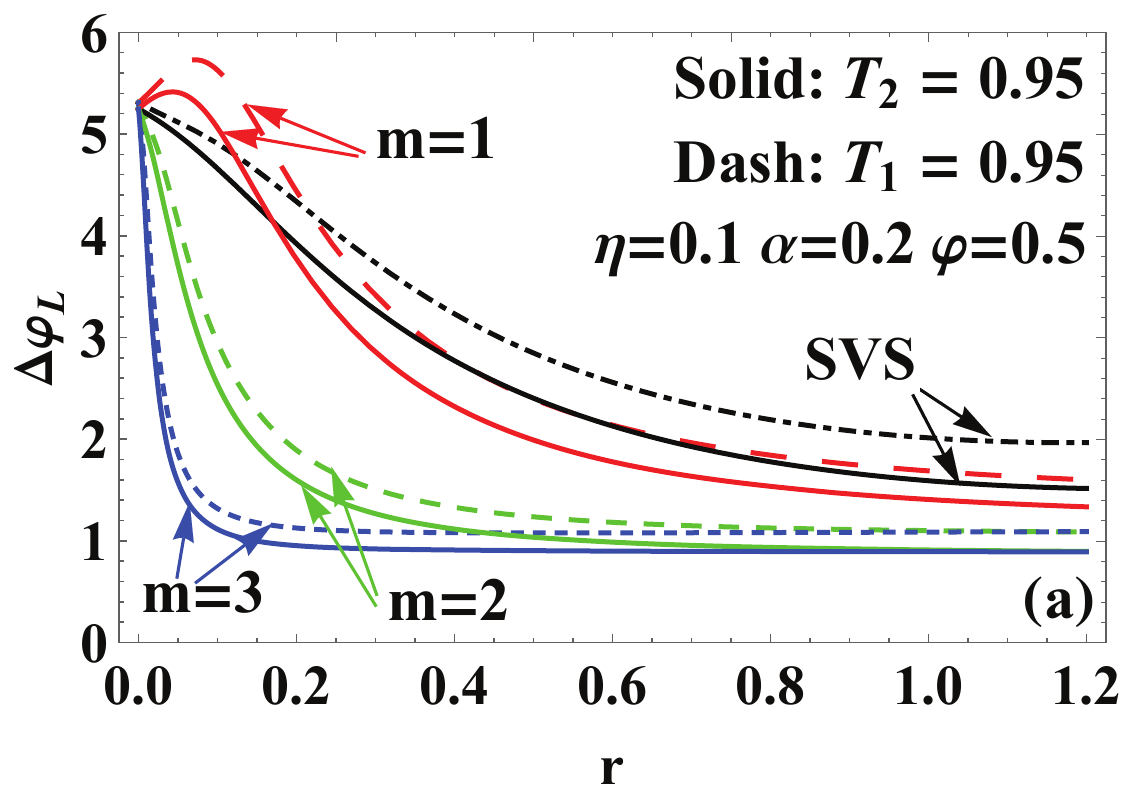}\\
\includegraphics[width=0.83\textwidth]{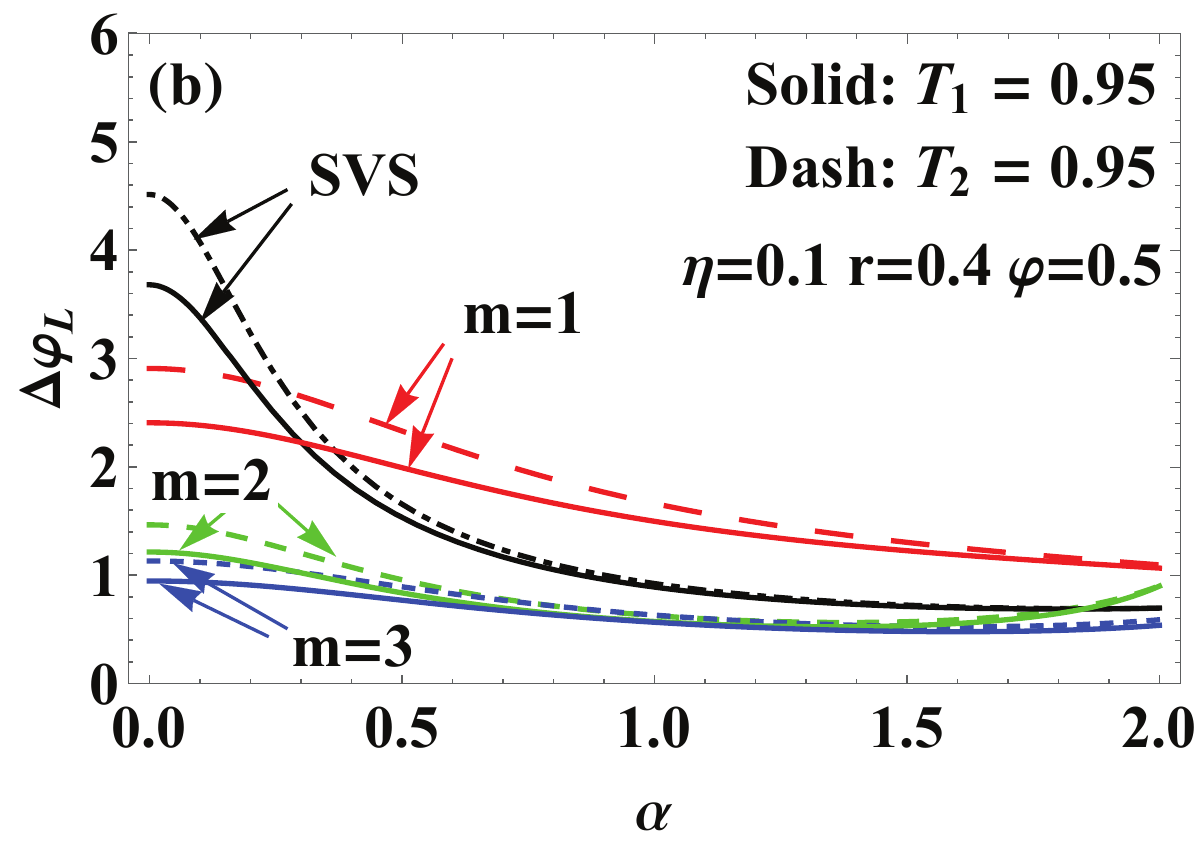}
\end{minipage}}
\caption{For the catalysis photon numbers $m=1,2,3$, and fixed phase shift $%
\protect \varphi =0.5$, with transmissivity values $T_{1}=T_{2}=0.95$ and $%
\protect \eta =0.1$, (a) the phase sensitivity $\Delta \protect \varphi _{L}$
under photon losses as a function of the squeezing parameter $r$ with the
coherent amplitude $\protect \alpha =0.2$, and (b) $\Delta \protect \varphi %
_{L}$ as a function of $\protect \alpha $ when $r=0.9$. The dashed (solid)
lines represent external (internal) dissipation, and the black lines
represent the SVS (without photon catalysis) for comparison.}
\end{figure}

From Fig. 12, the variation of the phase sensitivity $\Delta \varphi _{L}$
under photon losses with respect to the squeezing parameter $r$ and the
coherent amplitude $\alpha $ is visually observed under the conditions of $%
\varphi =0.5$, transmissivity $T_{1}=T_{2}=0.95$, and $\eta =0.1$. As
depicted in Fig. 12(a), it is evident that for $\alpha =0.2$, $\Delta
\varphi _{L}$ significantly improves with the increase of $r$ within a
specific range. Similarly, as indicated in Fig. 12(b), for the given
condition of $r=0.4$, $\Delta \varphi _{L}$ can improve with the increase of
$\alpha $ within a certain range. It is apparent from Fig. 12 that external
dissipation has a more pronounced effect on phase sensitivity. Furthermore,
in the presence of photon losses, the CS, when mixed with the PCSVS, can
notably enhance the phase sensitivity, particularly for the MC-SVS ($m=2,3$%
), relative to the SVS (without photon catalysis).

\section{\protect \bigskip The impact of photon losses on QFI}

\begin{figure}[tbh]
\label{Fig13} \centering \includegraphics[width=0.83\columnwidth]{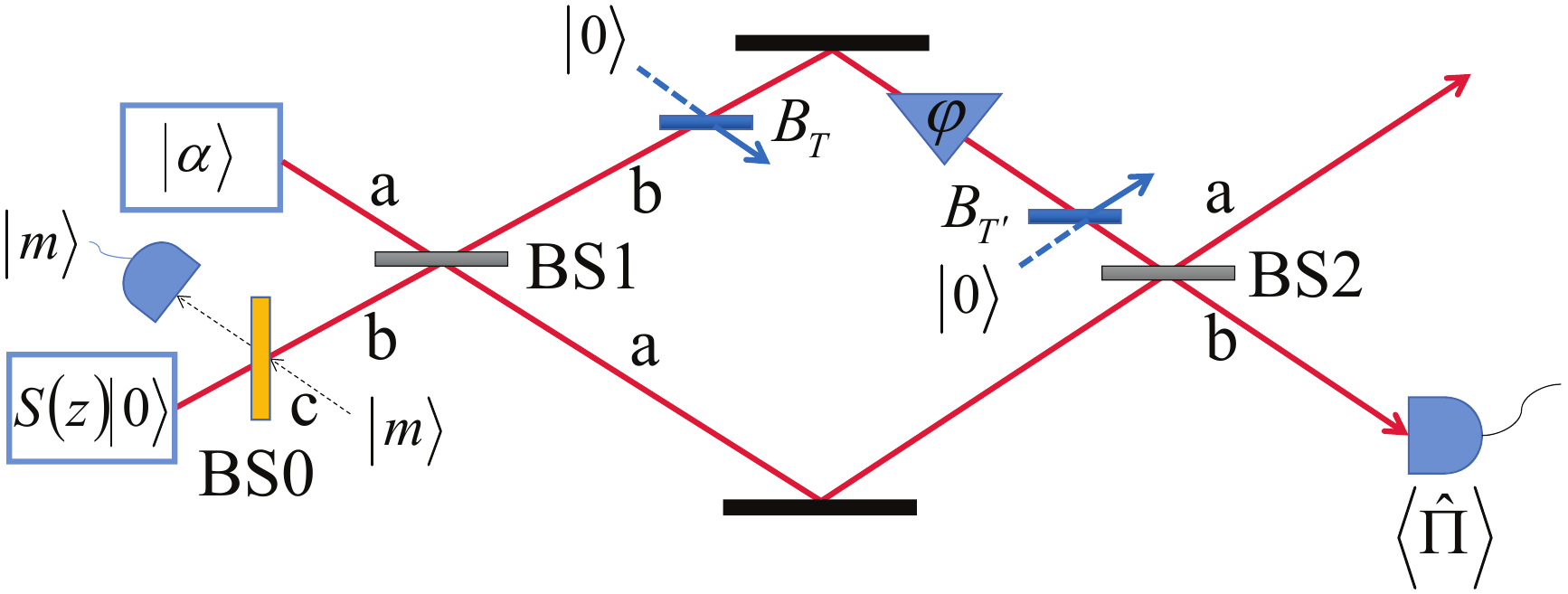}
\caption{Schematic diagram of lossy MZI with CS mixed PCSVS as input.}
\end{figure}

In the field of quantum metrology, the Cram\'{e}r-Rao bound theory of QFI
sets the limit for phase estimation. However, this limitation can be
influenced by environmental factors. This section focuses primarily on the
impact of photon losses in MZI on QFI. As depicted in Fig. 13, it is assumed
that photon losses occur in the optical path of mode $b$, primarily located
before and after the phase shifter, simulated by the optical BSs $B_{T}$ and
$B_{T^{^{\prime }}}$. According to Ref. \cite{54} in this scenario, the QFI
can be computed as

\begin{equation}
F_{QL}\leq C_{Q}=4\left[ \left \langle \psi \right \vert \hat{H}_{1}\left
\vert \psi \right \rangle -\left \vert \left \langle \psi \right \vert \hat{H%
}_{2}\left \vert \psi \right \rangle \right \vert ^{2}\right] ,  \label{37}
\end{equation}%
where, $\left \vert \psi \right \rangle =e^{-i\frac{\pi }{2}%
J_{1}}\left
\vert in\right \rangle $ is the quantum state that is input as
a pure state and passing through BS1. By utilizing Kraus operators
representation, we can express $H_{1,2}$ in the above equation as follows

\begin{eqnarray}
\hat{H}_{1} &=&\sum \limits_{l}\frac{d\hat{\Pi}_{l}^{\dagger }\left( \varphi
\right) }{d\varphi }\frac{d\hat{\Pi}_{l}\left( \varphi \right) }{d\varphi },
\notag \\
\hat{H}_{2} &=&i\sum \limits_{l}\frac{d\hat{\Pi}_{l}^{\dagger }\left(
\varphi \right) }{d\varphi }\hat{\Pi}_{l}\left( \varphi \right) ,  \label{38}
\end{eqnarray}%
where $\Pi _{l}\left( \varphi \right) $ is the Kraus operator

\begin{equation}
\hat{\Pi}_{l}\left( \varphi \right) =\sqrt{\frac{\left( 1-T\right) ^{l}}{l!}}%
e^{i\varphi \left( b^{\dagger }b-\gamma l\right) }T^{\frac{b^{\dagger }b}{2}%
}b^{l}.  \label{39}
\end{equation}%
In the equation, the transmissivity $T$ of optical BSs $B_{T}$ and $%
B_{T^{^{\prime }}}$ can be used to quantify the photon losses that occur on
mode $b$ in MZI. Here, $T=0$ and $T=1$ respectively represent complete
absorption and lossless scenarios. Parameter $\gamma =0$ and $\gamma =-1$
correspond to photon losses occurring before and after the phase shifter. By
optimizing $\gamma $, the minimum value of $C_{Q}$ can be achieved. Thus,
the QFI under the condition of photon losses can be obtained as

\begin{equation}
F_{QL}=\frac{4F_{Q}T\left \langle \psi \right \vert b^{\dagger }b\left \vert
\psi \right \rangle }{\left( 1-T\right) F_{Q}+4T\left \langle \psi \right
\vert b^{\dagger }b\left \vert \psi \right \rangle }.  \label{40}
\end{equation}%
By using unitary transformations, one can derive

\begin{equation}
\left \langle \psi \right \vert b^{\dagger }b\left \vert \psi \right \rangle
=\frac{1}{2}\left( \alpha ^{2}+\hat{D}\frac{\varepsilon }{\left(
1-4W_{1}W\right) ^{\frac{3}{2}}}-1\right) =\frac{1}{2}\bar{N}.  \label{41}
\end{equation}%
By substituting Eqs. (\ref{21}) and (\ref{41}) into Eq. (\ref{40}), one can
obtain $F_{QL}$.

\begin{figure}[tbh]
\label{Fig14} \centering%
\subfigure{
\begin{minipage}[b]{0.5\textwidth}
\includegraphics[width=0.83\textwidth]{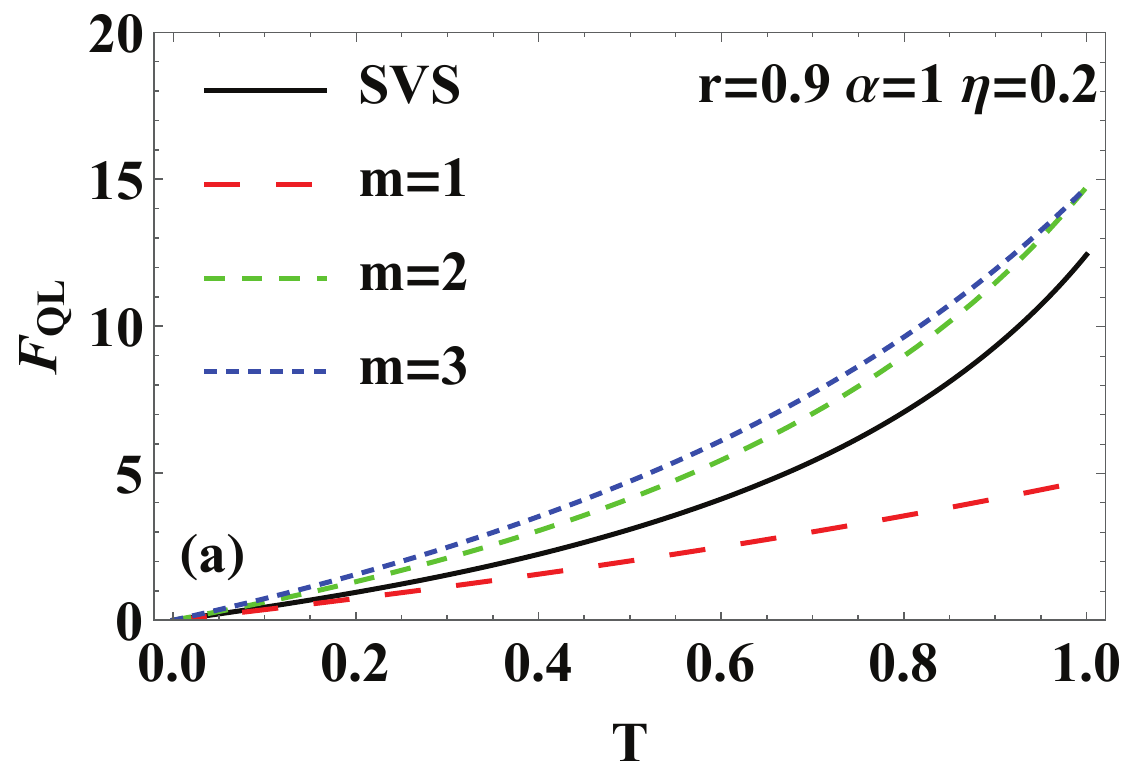}\\
\includegraphics[width=0.83\textwidth]{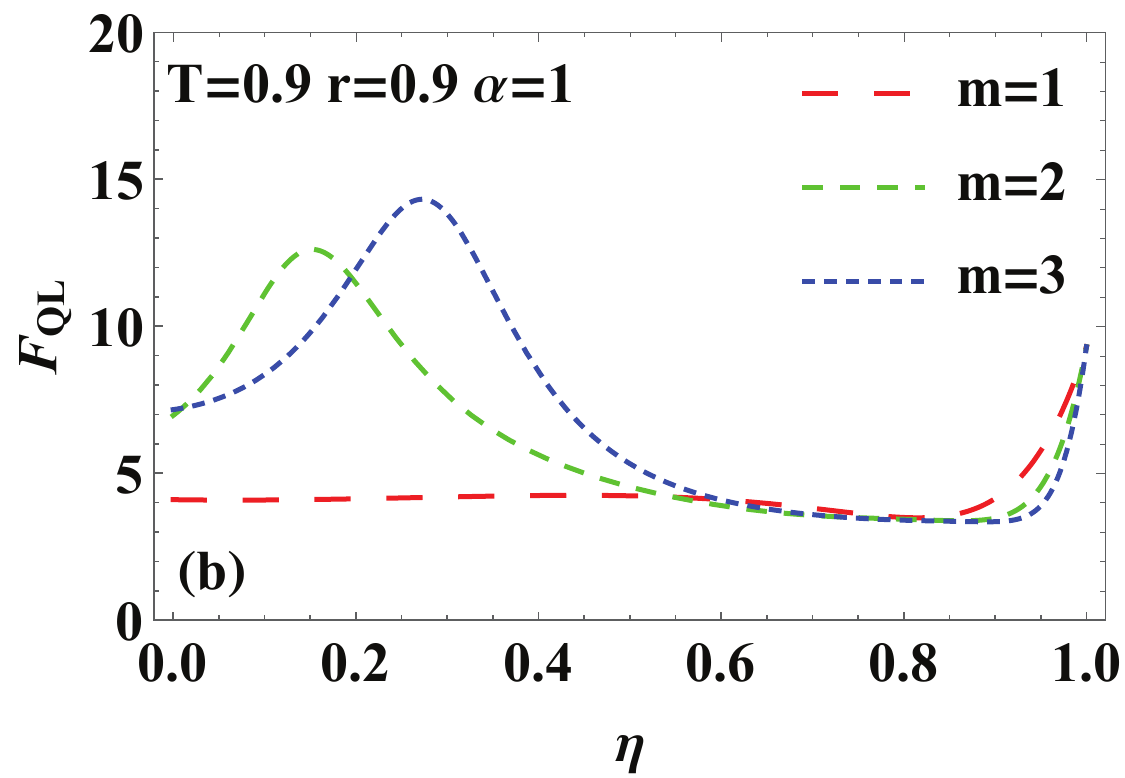}
\end{minipage}}
\caption{For the catalysis photon numbers $m=1,2,3$, fixed the squeezing
parameter $r=0.9$, and coherent amplitude $\protect \alpha =1$, (a) the
variation of $F_{QL}$ as a function of the transmissivity $T$ when $\protect%
\eta =0.2$, (b) $F_{QL}$ as a function of $\protect \eta $ when $T=0.9$. The
black solid line corresponds to the SVS (without photon catalysis) for
comparison.}
\end{figure}

In order to analyze the impact of photon losses on QFI, we can obtain the
variation of $F_{QL}$ with respect to its corresponding parameters based on
its analytical expression. Fig. 14(a) displays the variation of $F_{QL}$
with the transmissivity $T$ of $B_{T}$ ($B_{T^{^{\prime }}}$), which
simulates photon losses, for a given squeezing parameter $r=0.9$, coherent
amplitude $\alpha =1$, and transmissivity $\eta =0.2$. Intuitively, it is
observed that $F_{QL}$ increases with $T$, indicating that the QFI rises as
photon losses decrease. Additionally, the QFI for the CS mixed with the
MC-SVS ($m=2,3$), can increase relative to the case of CS mixed with the SVS
as inputs. Furthermore, from Fig. 14(b), it is evident that the change in $%
F_{QL}$ with $\eta $, for $T=0.9$ with photon losses, when $\alpha =1$ and $%
r=0.9$, the case of $m=2,3$ can significantly increase compared to the SVS
(without photon catalysis) case when $\eta $ is small.
\begin{figure}[tbh]
\label{Fig15} \centering%
\subfigure{
\begin{minipage}[b]{0.5\textwidth}
\includegraphics[width=0.83\textwidth]{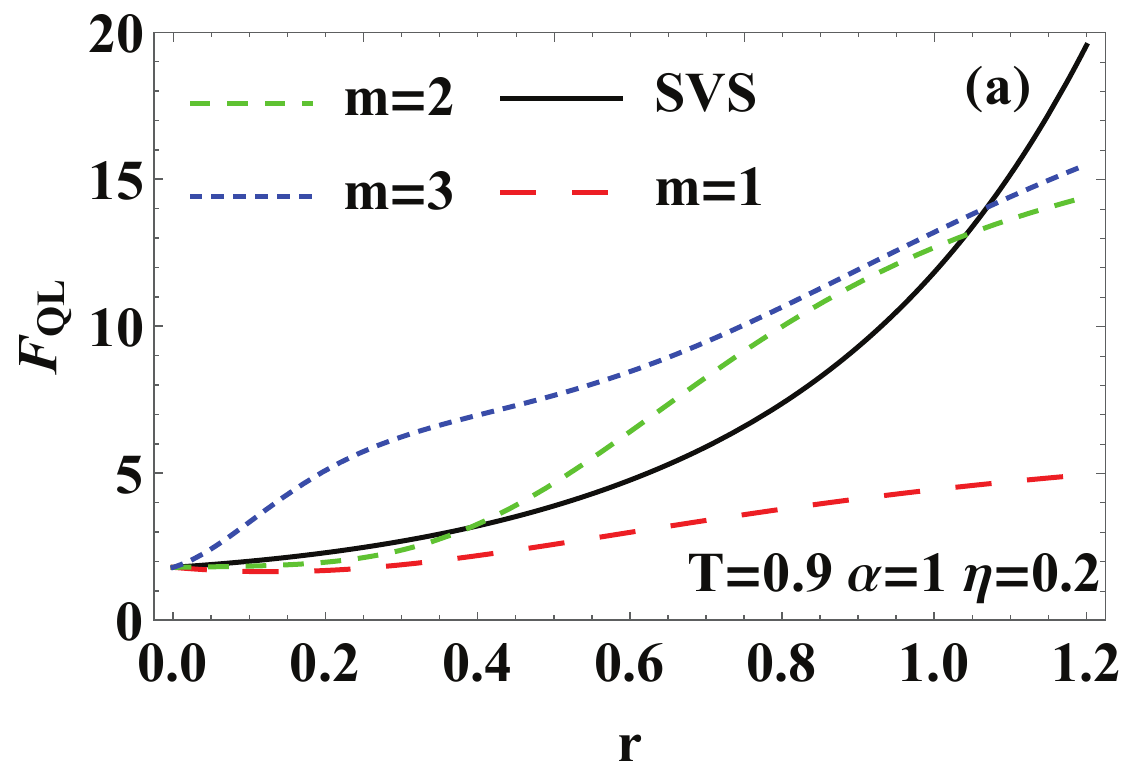}\\
\includegraphics[width=0.83\textwidth]{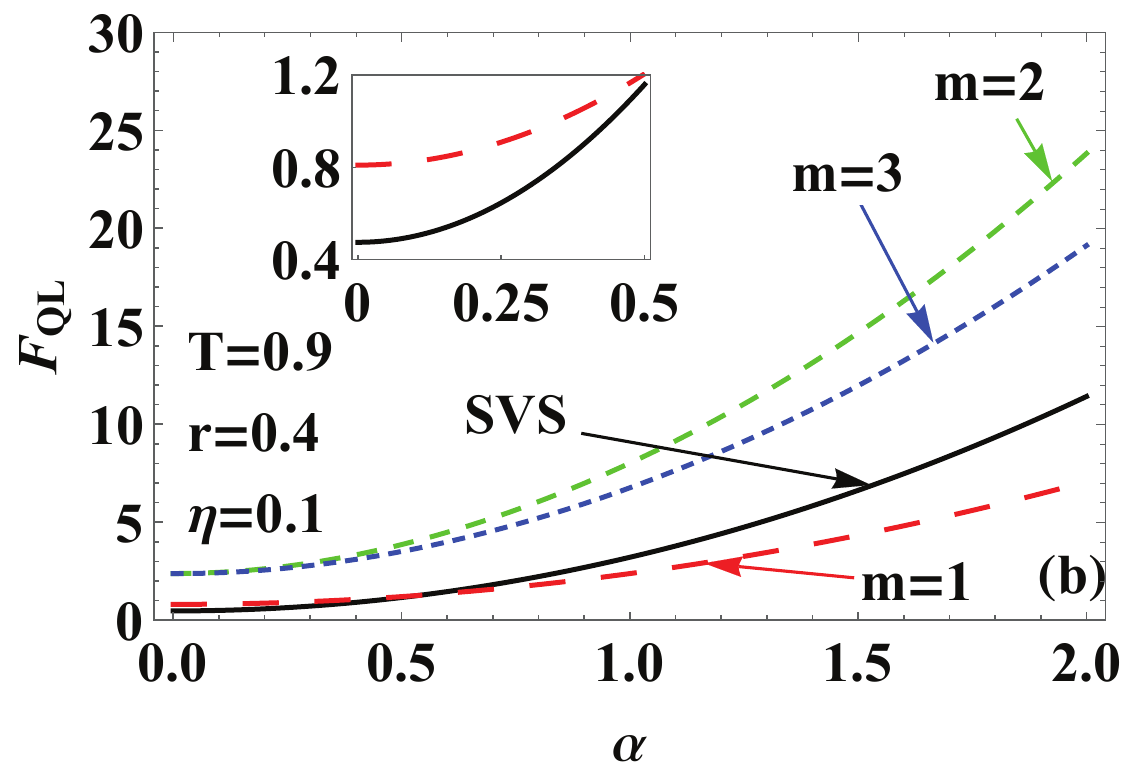}\\
\includegraphics[width=0.83\textwidth]{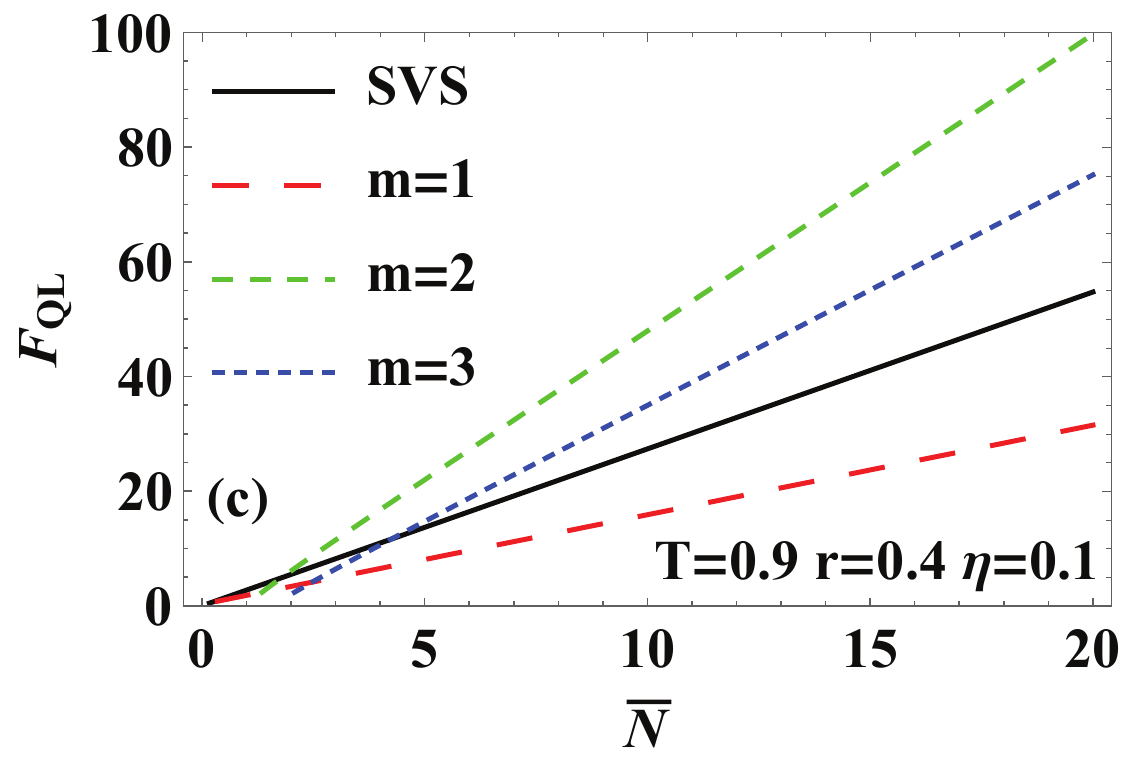}
\end{minipage}}
\caption{For the catalysis photon numbers $m=1,2,3$, and fixed the
transmissivity $T=0.9$, (a) $F_{QL}$ as a function of the squeezing
parameter $r$ when the coherent amplitude $\protect \alpha =1$ and $\protect%
\eta =0.2$, (b) $F_{QL}$ as a function of $\protect \alpha $ when $r=0.4$ and
$\protect \eta =0.1$, and (c) $F_{QL}$ as a function of average photon number
$\bar{N}$ when $r=0.4$ and $\protect \eta =0.1$. The black solid line
corresponds to the SVS (without photon catalysis) for comparison.}
\end{figure}

Fig. 15 illustrates the intuitive analysis of the relationship between $%
F_{QL}$ and the relevant parameters of the input resources in the presence
of photon losses ($T=0.9$) with a fixed small transmissivity $\eta $. The
clear variation pattern in Fig. 15 indicates that, even with photon losses,
the QFI can increase by raising the squeezing parameter $r$, coherent
amplitude $\alpha $, and average photon number $\bar{N}$. Additionally, it
demonstrates that the CS mixed with MC-SVS with $m=2,3$ as input states can
effectively increase the $F_{QL}$ compared to the input CS mixed with SVS.
Particularly, from Fig. 15(b), it is evident that for the case of $m=1$ and
a relatively small $r=0.4$, $F_{QL}$ can also be increased relative to SVS
(without photon catalysis) in the range of small $\alpha $, which is
relatively easy to achieve experimentally.

\begin{figure}[tbh]
\label{Fig16} \centering%
\subfigure{
\begin{minipage}[b]{0.5\textwidth}
\includegraphics[width=0.83\textwidth]{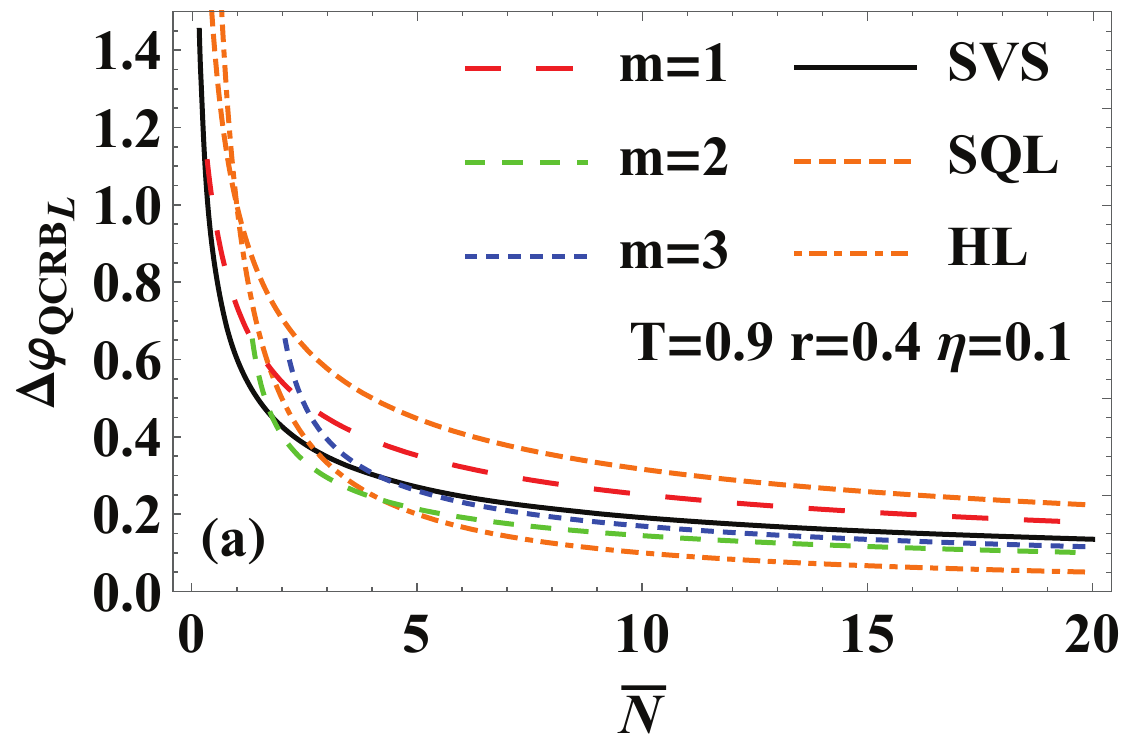}\\
\includegraphics[width=0.83\textwidth]{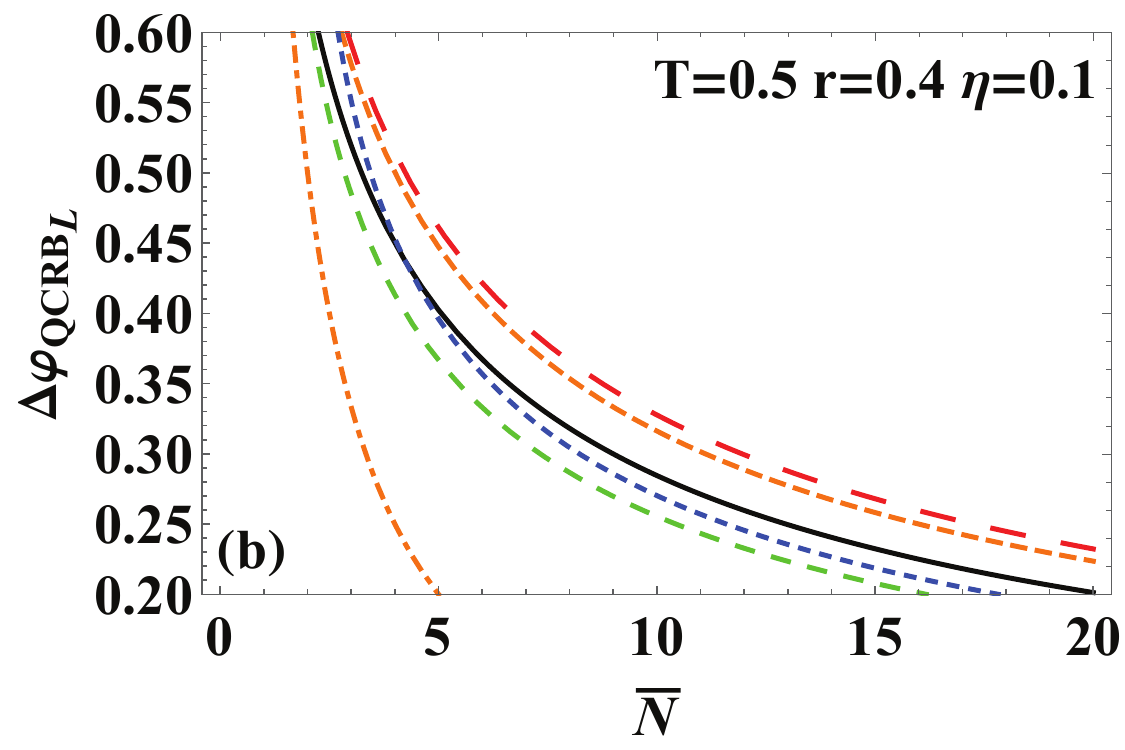}
\end{minipage}}
\caption{For the catalysis photon numbers $m=1,2,3$, and fixed the squeezing
parameter $r=0.4$, and transmissivity $\protect \eta =0.1$, (a) with
transmissivity $T=0.9$, (b) with $T=0.5$, the variation of QCRB $\Delta
\protect \varphi _{QCRB_{L}}$ with respect to the average photon number $%
\overline{N}$. The black solid line, orange dashed line, and orange
dotdashed line is respectively represent the SVS (without photon catalysis),
SQL, and HL for comparison.}
\end{figure}

The variation of the QCRB ($\Delta \varphi _{QCRB_{L}}=1/\sqrt{F_{QL}}$)
with the energy of the input CS mixed with PCSVS (corresponding to the
average photon number $\bar{N}$) under the condition of photon losses, and
its comparison with the case where CS mixed with SVS is used as the input
state, are depicted in Fig. 16. The SQL and the HL are also included for
comparison. The figure illustrates an improvement in QCRB with the increase
of $\bar{N}$, and it exhibits a certain enhancement relative to CS mixed
with SVS when the CS mixed with MC-SVS ($m=2,3$) as inputs. Furthermore,
Fig. 16(a) clearly demonstrates that despite the existence of photon losses
corresponding to transmissivity $T=0.9$, the QCRB can effectively surpass
the SQL. Moreover, for small $\bar{N}$ values, the QCRB can approach or even
exceed the HL. Additionally, as depicted in Fig. 16(b), it is observed that
the case of $m=2,3$ can successfully exceed the SQL, even for relatively
high photon losses corresponding to $T=0.5$. This observation indicates the
strong robustness of the QCRB for the scheme involving the CS mixed with
MC-SVS as inputs against photon losses.

\section{\protect \bigskip Conclusion}

In this study, we focused on investigating the precision of phase
measurement in MZI using a CS mixed with PCSVS as input state, employing
both the phase sensitivity of parity detection and the QFI. The findings
indicate that the incorporation of photon catalysis, particularly through
multi-photon catalysis operations, leads to a substantial enhancement in
phase measurement precision compared to using CS mixed with SVS as input. In
an ideal scenario involving the input of MZI with CS mixed with PCSVS, and
with the implementation of parity detection at the output port, the
adjustment of the transmissivity $\eta $ of the BS preparing the PCSVS
can effectively enhance the resolution of the phase shift $\varphi $.
Consequently, the optimization of $\eta $ is crucial for improving phase
sensitivity. It was observed that the optimal point for $\Delta \varphi $
occurs at $\varphi =0$, and for relatively small $\eta $, the phase
sensitivity $\Delta \varphi $ and QFI are significantly improved in the
scheme of employing CS mixed with MC-SVS (the catalysis photon numbers $%
m=2,3 $) as input in MZI, as compared to CS mixed with SVS as the input
state. Furthermore, when the squeezing parameter $r$ and coherent amplitude $%
\alpha $ are small, the phase sensitivity $\Delta \varphi $ and QFI of $m=1$
is also enhanced relative to that of the SVS (without photon catalysis).
Moreover, by increasing $r$, $\alpha $ and the total average number of
photons $\overline{N}$ corresponding to the energy of the input resource,
the QFI and $\Delta \varphi $ can be improved. Notably, in the case of $%
m=2,3 $, the phase sensitivity $\Delta \varphi $ and the QCRB can surpass
the SQL and even exceed the HL when compared with CS mixed with SVS as the
input state.

Our study delved further into analyzing the phase sensitivity and QFI in the
presence of photon losses in practical scenarios. The findings revealed that
external dissipation has a more pronounced impact on phase sensitivity than
internal dissipation. Although photon losses will reduce phase sensitivity
and QFI, it is still possible to increase $r$, $\alpha $, and $\overline{N}$
at low $\eta $ to make multi-photon catalysis operation more effectively
improve phase sensitivity and QFI. Additionally, the mixing of CS and MC-SVS
can significantly improve phase sensitivity and the QFI better than the case
of CS mixed with SVS as the input state, while there is also a certain
improvement for $m=1$ when $r$ and $\alpha $ are relatively small. Moreover,
in the presence of photon losses, the QCRB can still significantly
outperform the SQL.

Overall, employing the scheme of mixing CS with PCSVS, especially when mixed
with MC-SVS as input in MZI, significantly enhances phase measurement
accuracy. Our findings hold significant importance in advancing quantum
measurement for practical applications.

\begin{acknowledgments}
This work is supported by the National Natural Science Foundation of China
(Grants No. 11964013 and No. 12104195) and the Training Program for Academic
and Technical Leaders of Major Disciplines in Jiangxi Province (No.
20204BCJL22053).
\end{acknowledgments}

\end{document}